\documentclass[lettersize,journal]{IEEEtran}
\IEEEoverridecommandlockouts
\usepackage{array}
\usepackage{url}
\usepackage{verbatim}
\usepackage{paralist}
\usepackage{graphicx}
\usepackage{color}
\usepackage{pifont}
\usepackage[linesnumbered,ruled,vlined]{algorithm2e}
\usepackage{comment}
\usepackage[bookmarks=false]{hyperref} 
\usepackage{cite}
\hypersetup{colorlinks=true,citecolor=blue}

\usepackage{amsmath,amssymb,amsfonts}
\usepackage{graphicx}
\usepackage{textcomp}
\usepackage{caption}
\usepackage{subcaption}

\usepackage{algpseudocode}
\usepackage{enumitem,lipsum}
\usepackage{csquotes}
\setlist[itemize,enumerate]{leftmargin=*}
\usepackage{etoolbox}
\makeatletter
\usepackage{booktabs,makecell,threeparttable,tabularx,array,wasysym}
\usepackage{multirow}
\usepackage{listings}
\newtheorem{definition}{Definition}[section]

\newcommand{\ie}{\textit{i.e.}}
\newcommand{\eg}{\textit{e.g.}}

\setlist[enumerate,1]{label=\textbf{\scriptsize\arabic*.}, leftmargin=*}

\newlist{circlelist}{enumerate}{1}
\setlist[circlelist,1]{label=\textbf{\textcircled{\small\arabic*}}, leftmargin=*}

\usepackage{tcolorbox}
\tcbuselibrary{breakable}
\usepackage{orcidlink}

\usetikzlibrary{decorations.pathreplacing}

\newcounter{protocol}
\renewcommand{\theprotocol}{\arabic{protocol}}

\newtcolorbox{protocolbox}[2][]{
	breakable,
	boxrule=1pt,
	colframe=gray!40,      
	colback=white,         
	coltitle=black,        
	title={\uppercase{Protocol}~\theprotocol: #2},
	#1                     
}

\begin{document}
	
	\title{$\mathsf{DARTIC}$: Decentralized Anonymous Reputation at Scale for Trustworthy  Crowdsourcing}

	\author{
		
		\IEEEauthorblockN{Mouhamed Amine Bouchiha, \textit{Member}, \textit{IEEE}, Mourad Rabah, Ronan Champagnat, Abdelaziz Amara Korba, \textit{Member}, \textit{IEEE}, Yacine Ghamri-Doudane,  \textit{Senior Member}, \textit{IEEE}} \\

		\thanks{M. A. Bouchiha is with SAMOVAR, Telecom-SudParis, Paris, France;  M. Rabah, R. Champagnat, and Y. Ghamri-Doudane are with L3i, La Rochelle University, France; A. Amara Korba is with the Department of Computer Science, German University of Technology in Oman (GUtech), Muscat, Oman; (e-mails: mbouchiha@telecom-sudparis.eu.fr, \{mourad.rabah, ronan.champagnat, yacine.ghamri\}@univ-lr.fr, abdelaziz.amarakorba@gutech.edu.om)}}

	\markboth{IEEE Transactions on Services Computing
		,~Vol.~XX, No.~X, XX~2026}%
	{Shell \MakeLowercase{\textit{et al.}}: A Sample Article Using IEEEtran.cls for IEEE Journals}
	
	\maketitle

	\begin{abstract}
		On-chain crowdsourcing leverages blockchain’s decentralization, transparency, and tamper-resistance to build trustworthy and verifiable Web3 crowdsourced services. However, existing decentralized reputation frameworks do not reconcile anonymity, reputation binding, and scalability. This paper demonstrates how on-chain crowdsourcing can simultaneously achieve these requirements under a trust-minimized model. We introduce $\mathsf{DARTIC}$, a decentralized, anonymous, and scalable reputation-driven framework for crowdsourcing. $\mathsf{DARTIC}$ presents a dual-ledger system that enables requesters and workers to use distinct pseudonyms across interactions, ensuring unlinkability while maintaining accountability. To mitigate Sybil and reputation-reset attacks, we employ zkSNARK-based set membership proofs, cryptographically binding all user pseudonyms to a single access token without revealing the linkage. For scalability, we investigate two aggregation techniques that compress multiple proofs into a single succinct proof to minimize verification overhead. In addition, we design an automated, privacy-preserving reputation model that dynamically evaluates contributions across diverse crowdsourcing contexts. To demonstrate practicality, we instantiate and assess $\mathsf{DARTIC}$ in both crowdsensing and federated learning scenarios. Experimental results show that (i) individual proof generation for token spending completes in less than $3$s, (ii) aggregation reduces the verification time of 1024 proofs from $8.7$s to $0.96$s, and (iii) zk-batching lowers gas costs by more than 100× compared to a pure Layer-1 deployment. These results demonstrate that anonymity, robust reputation binding, and scalability can be jointly achieved in fully decentralized crowdsourcing systems.
	\end{abstract}
	
	
	\section{Introduction}
	\quad 
	
	\IEEEPARstart{W}{eb3}-enabled crowdsourcing~\cite{b9, trustworker} is an emerging class of \emph{decentralized service computing systems}~\cite{liu2019decentralized}, where a blockchain network coordinates, validates, and enforces service interactions between requesters and providers (aka workers) without relying on centralized intermediaries. Unlike traditional crowdsourcing platforms, where service discovery, execution, and trust management are centrally controlled, Web3-based models decentralize service orchestration, enhancing transparency, security, and fault tolerance~\cite{b9, zebralancer, bcmcs}. In this paradigm, crowdsourcing tasks can be modeled as \emph{service provisioning workflows}, where workers offer on-demand services and requesters consume them under quality-of-service expectations.
	
	\quad A fundamental challenge in such decentralized ecosystems is \emph{trust establishment} among mutually distrustful participants. Reputation-driven blockchain-based crowdsourcing~\cite{crowdr, pprep} has emerged as a promising approach to support trust-aware task allocation and incentive-compatible service execution. In these systems, workers—and in some cases requesters—are assigned reputation scores that influence task assignment, pricing, and rewards~\cite{rmanager,acmnetcrowd24}. Reputation acts as a soft-state quality indicator complementing functional service descriptions~\cite{repcn}. However, existing frameworks remain vulnerable to strategic manipulation, including whitewashing (aka reset), self-promotion, and bad-collision attacks~\cite{pprep}. To mitigate such threats, many solutions bind reputation to static identifiers such as addresses, certificates, or public keys. While effective for accountability, this approach undermines privacy and discourages honest feedback, as participants may fear retaliation, de-anonymization, or long-term profiling.
	
	\quad Blockchain-based privacy-preserving reputation mechanisms enhance decentralized trust by protecting \emph{reputation data privacy} (e.g., ratings and computation inputs)~\cite{pprep}. However, \emph{user privacy} remains largely unaddressed. Due to blockchain’s transparency, on-chain interactions can reveal behavioral patterns and transaction links, enabling traceability and potential deanonymization even when reputation values are hidden. In crowdsensing, such visibility may also encourage free-riding~\cite{huang2022blocksense}, while public feedback can expose evaluators to retaliation. Although automated, feedback-free models~\cite{b9,autodfl} reduce direct exposure of ratings, binding reputation to persistent identifiers or master keys still permits long-term tracking. Integrating cryptographic primitives such as zero-knowledge proofs with blockchain platforms~\cite{arsps, dars} offers a promising avenue for building systems that balance privacy and accountability. Yet, current frameworks remain limited. 
	Many solutions~\cite{bcmcs, trustworker, anoTFPP} rely on centralized authorities to manage identity credentials, thus introducing trusted intermediaries and preventing a fully trust-minimized design. Others leverage privacy-enhancing blockchains~\cite{drep,trustpriv,brpp2024}, but do not simultaneously address scalability, cost-efficiency, and automated evaluation with secure integration of real-world service data. These gaps raise the following central research question.
	
	\begin{list}{}{\leftmargin=1em \rightmargin=1em}
		\item \textit{How can decentralized crowdsourcing reconcile anonymity, reputation binding, and scalability without relying on trusted authorities?}
	\end{list}
	
	\begin{table*}[th]
		\centering
		\caption{$\mathsf{DARTIC}$ vs. related privacy-preserving/anonymous reputation systems.}
		\label{tab:differences}
		\setlength{\tabcolsep}{2.5pt}
		\scalebox{0.99}{
			\begin{threeparttable}
				\begin{tabular}{ccccccccccc}
					
					\multicolumn{1}{c}{\rotatebox{30}{Reference}} & 
					\multicolumn{1}{c}{\rotatebox{30}{\parbox{4.5em}{\centering \footnotesize Threat Model}}} & 
					\multicolumn{1}{c}{\rotatebox{30}{\parbox{4.5em}{\centering \footnotesize Collusion Resistance}}} & 
					\rotatebox{30}{\parbox{4.5em}{\centering \footnotesize Trustless}} & 
					\multicolumn{1}{c}{\rotatebox{30}{\parbox{4.5em}{\centering \footnotesize Reputation Binding}}} & 
					\multicolumn{1}{c}{\rotatebox{30}{\parbox{5em}{\centering \footnotesize Reputation Aggregation}}} & 
					\multicolumn{1}{c}{\rotatebox{30}{\parbox{4.5em}{\centering \footnotesize Multi-Pseud}}} & 
					\multicolumn{1}{c}{\rotatebox{30}{\parbox{4.5em}{\centering \footnotesize (Rater/Ratee) Anonymity}}} & 
					\rotatebox{30}{\parbox{4.5em}{\centering \footnotesize Accountability}} & 
					\multicolumn{1}{c}{\rotatebox{30}{\parbox{5em}{\centering \footnotesize Reputation Automation}}} &
					\rotatebox{30}{\parbox{5em}{\centering \footnotesize Scalability}} \\
					\hline
					Gao et al. \cite{trustworker} & S-H & \LEFTcircle & \Circle & ID & W-Sum & \Circle & \Circle & \CIRCLE & \Circle & \Circle \\
					Zhao et al. \cite{bcmcs} & S-H & \LEFTcircle & \Circle & ID & Mean & \Circle & \Circle & \CIRCLE & \Circle & \Circle \\
					Liu et al. \cite{arsps} & Mal. & \CIRCLE & \Circle & ID & Sum & \Circle & \LEFTcircle & \CIRCLE & \Circle & \Circle \\
					Dimitriou \cite{drep} & S-H/Mal. & \CIRCLE & \LEFTcircle & ID & Sum & \CIRCLE & \CIRCLE & \LEFTcircle & \Circle & \Circle \\
					Deng et al. \cite{brpp2024} & S-H/Mal. & \CIRCLE & \LEFTcircle & Pseudo & W-Mean & \CIRCLE & \CIRCLE & \LEFTcircle & \Circle & \LEFTcircle \\
					Chen et al. \cite{anoTFPP} & S-H & \LEFTcircle & \Circle & Pseudo & W-Sum & \CIRCLE & \CIRCLE & \LEFTcircle & \Circle & \CIRCLE \tnote{*} \\
					Duan et al. \cite{duan2019aggregating}  & S-H/Mal. & \LEFTcircle  & \LEFTcircle  & ID  & Sum+DP  & \Circle  & \LEFTcircle & \CIRCLE  & \LEFTcircle & \LEFTcircle \\
					Koutsos et al. \cite{koutsos2024mathsf} & S-H/Mal. & \LEFTcircle & \Circle & ID & Sum & \CIRCLE & \CIRCLE & \CIRCLE & \Circle & \CIRCLE \\
					\textbf{$\mathsf{DARTIC}$} & Mal. & \CIRCLE & \CIRCLE & Pseudo & PW-Mean & \CIRCLE & \CIRCLE & \CIRCLE & \CIRCLE & \CIRCLE \\
					\hline
				\end{tabular}
				\begin{tablenotes}
					\footnotesize
					\item Symbols: \CIRCLE = Full support; \LEFTcircle = Partial; \Circle = No support.  
					Mal. = Malicious; S-H = Semi-honest; PW = Piecewise Weighted;  DP = Differential Privacy.  
					\item[] $^*$Achieves scalability but remains centralized.  
					$\mathsf{DARTIC}$ uniquely combines \textit{scalable trustless decentralization} and \textit{fully automated reputation}, with strong anonymity guarantees for both raters and ratees.
				\end{tablenotes}
			\end{threeparttable}
		}
	\end{table*}

	\quad Addressing this question requires overcoming three fundamental tensions: 
	(i) anonymity \emph{vs.} accountability, 
	(ii) decentralization \emph{vs.} efficient identity management, and 
	(iii) privacy-preserving verification \emph{vs.} on-chain scalability. 
	
	\quad To resolve these tensions, we propose $\mathsf{DARTIC}$, a decentralized, anonymous, and reputation-driven crowdsourcing framework.  Table~\ref{tab:differences} positions $\mathsf{DARTIC}$ with respect to prior work. Unlike existing solutions, $\mathsf{DARTIC}$ jointly achieves: \begin{inparaenum}[(i)] \item trustless reputation binding without persistent identifiers, \item fully decentralized and automated reputation, and \item scalable on-chain verification. \end{inparaenum}The main contributions of this work are summarized as follows:
	\begin{itemize}
		\item \textit{Decoupled identity and service management for trustless reputation binding.} 
		We design a dual-ledger architecture that separates identity control from service interactions, enabling privacy-preserving execution while ensuring Sybil resistance via zkSNARK proofs, decentralized oracles, and threshold cryptography, eliminating reliance on trusted authorities.
		\item \textit{Unlinkable pseudonymous participation with contextual accountability.} 
		$\mathsf{DARTIC}$ allows requesters and workers to generate multiple pseudonyms while cryptographically binding them to a single access token. This design preserves unlinkability across interactions while preventing whitewashing.
		\item \textit{Automated and privacy-preserving reputation evaluation.} 
		We introduce a general reputation model driven by verifiable service execution outcomes rather than explicit feedback. This reduces  retaliation and manipulation risks while maintaining accountability and resistance to strategic behavior.
		\item \textit{Scalable verification through proof aggregation and Layer-2 execution.} 
		To ensure practical deployability, we integrate batching and aggregation mechanisms, significantly reducing on-chain verification overhead and transaction costs while preserving security guarantees.
	\end{itemize}
	
	The remainder of this paper is organized as follows. \S\ref{sec:$DARTC$relatedworks} reviews related work. \S\ref{sec:$DARTC$Arch} presents the proposed $\mathsf{DARTIC}$ framework. \S\ref{sec:$DARTC$math} details its protocols. \S\ref{sec:$DARTC$security} provides theoretical analysis, while \S\ref{sec:$DARTC$evaluation} discusses performance evaluation. Finally, \S\ref{sec:$DARTC$conclusion} concludes the paper.

	\section{Related Work} \label{sec:$DARTC$relatedworks}
	
	\noindent\textit{Blockchain-Based Reputation Systems with Centralized or Semi-Trusted Authorities.}  
	Several works employ blockchain to ensure transparency and verifiability while relying on semi-trusted or centralized entities for identity and credential management.  
	Liu et al.~\cite{arsps} present an anonymous reputation system for industrial retail IoT marketing, using PoS-enabled smart contracts, randomizable signatures, and non-interactive zero-knowledge proofs to guarantee anonymity. However, the Identity Management (IDM) system presents a single point of failure and a trust issue.  
	Similarly, Zhao et al.~\cite{bcmcs} propose a privacy-preserving blockchain-based mobile crowdsensing reputation framework using additive secret sharing and delegation sets. While smart contracts update global scores transparently, the model assumes a semi-honest Task Distribution Center (TDC) and a fully trusted Key Distribution Center (KDC).  
	Chen et al.~\cite{anoTFPP} adopt a three-layer architecture (task publishing, reputation management, computation) with blind and group signatures to anonymize workers, and a Shapley value-based mechanism for fair reward distribution. Despite advanced defenses against malicious workers and collusion, reliance on an honest but curious model restricts applicability.  
	
	\medskip
	\noindent\textit{Lightweight Privacy-Preserving Approaches in Crowdsourcing and Crowdsensing.}  
	Focusing on efficiency in constrained environments, Deng et al.~\cite{brpp2024} propose a blockchain-based recruitment scheme using Pedersen commitments and CLSAG ring signatures. While lightweight and privacy-preserving, it assumes unbiased requester evaluations and overlooks systemic barriers to consistent worker participation.  
	TrustWorker~\cite{trustworker} introduces a secure two-party protocol to compare worker reputations without disclosure, offering privacy with low overhead. Nonetheless, it ignores requester reputations and assumes semi-honest adversaries with static ID bindings, potentially reducing adaptability in dynamic ecosystems.  
	
	\medskip
	\noindent\textit{Advanced Cryptographic Integration for Privacy and Verifiability.}  
	FedCrowd~\cite{guo2020fedcrowd} presents an innovative integration of federated crowdsourcing with blockchain for privacy-preserving task recommendation. While privacy is addressed through encrypted task queries, the model primarily focuses on protecting task attributes rather than ensuring the long-term anonymity and unlinkability of workers. AVeCQ~\cite{koutsos2024mathsf} exemplifies the integration of advanced cryptography into blockchain-based crowdsourcing. Built on Ethereum smart contracts, it combines zkSNARKs for verifiable quality proofs, Pedersen commitments for homomorphic updates, and Merkle trees for membership verification. Its semi-honest Registration Authority (RA) balances decentralization with efficiency, but trust poses a potential risk.  
	
	\medskip
	\quad Across these systems and others~\cite{b19, copifl, zebralancer, duan2019aggregating}, blockchain serves as a verifiable, tamper-resistant ledger, and privacy is enforced through signatures, commitments, encryption, or zero-knowledge proofs. However, most frameworks retain \emph{semi-trusted or centralized authorities} (\eg, IDM, KDC, TPC, RA), posing trust and centralization issues. Furthermore, many assume \emph{benign evaluator behavior} and \emph{semi-honest adversaries}, assumptions that may not hold in practice when facing collusion, biased evaluations, or reputation attacks.
	
	\quad We address these gaps in this work by introducing $\mathsf{DARTIC}$, a fully decentralized, anonymous, and reputation-driven framework for crowdsourcing. To achieve end-to-end decentralization and secure integration of real-world data, $\mathsf{DARTIC}$ leverages a dual-ledger architecture alongside a decentralized oracle network, supported by a privacy-preserving protocol~\cite{deco}. The system enables users to interact anonymously under multiple pseudonyms while maintaining a robust reputation through Merkle trees and scalable zkSNARK constructions. Combined with an automated evaluation, this design ensures that honest users retain their earned reputation by mitigating Sybil and reputational attacks (\eg, bad-mouthing and self-promotion). Collectively, these features establish $\mathsf{DARTIC}$ as a robust and practical solution for decentralized, anonymous, and verifiable crowdsourcing in real-world settings.
	
	\begin{table}[t]
		\centering
		\caption{ Summary of key notations used in \(\mathsf{DARTIC}\).}
		\setlength{\tabcolsep}{3pt}
		
		\begin{tabular}{m{3.0cm} m{5.0cm}}
			\toprule
			\textbf{Notation} & \textbf{Description} \\
			\midrule
			$AT$ & Access Token \\
			$RT$ & Reputation Token \\
			$u, v$ & Users participating in crowdsourcing \\
			$CS = \{ cn_i \}$ & Set of IDML committee nodes \\
			$M_{pk}, M_{sk}$ & Master public and private keys of a user \\
			$M_{cred}$ & Master credential issued by IDML \\
			$\sigma_{cred}$ & Committee signature on $M_{cred}$ \\
			$ctx$ & Application-specific context \\
			$(apk_u, ask_u)$ & Address public/private key pair \\
			$R_u$  & Reputation score of user $u$ \\
			$\mathsf{W_f}$ & Interaction weighting function \\
			$T^{v}$ & Contribution from a worker $v$ \\
			$cm_A$ & Commitment to an access credential \\
			$cm_R$ & Commitment to a reputation credential \\
			$ACTree$ & Merkle tree of access commitments \\
			$RCTree$ & Merkle tree of reputation  commitments \\
			$DCTree$ & Merkle tree of deposit  commitments \\
			$\mathsf{COMM}(\cdot)$ & Cryptographic commitment function \\
			$pk_{cs}$ & Public key of IDML committee \\
			$sk_{cs}$ & Secret signing key of IDML committee \\
			$\sigma_{orc}$ & Threshold signature generated by DON \\

			\bottomrule
		\end{tabular}
		\label{tab:notations_dartic}
	\end{table}

	\section{$\mathsf{DARTIC}$ Framework} \label{sec:$DARTC$Arch} 
	
\begin{figure}[t]
		\centering
		\includegraphics[width=1\linewidth, height=0.6\linewidth]{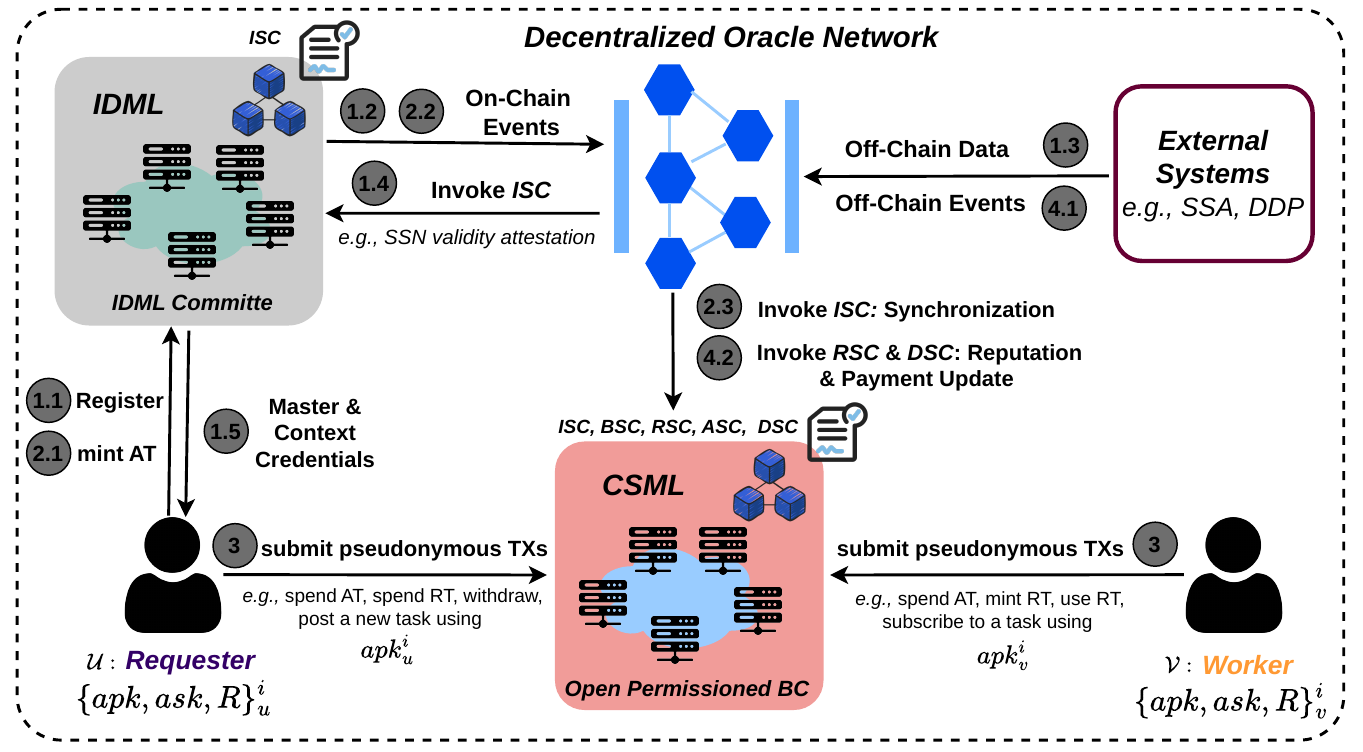}
		\caption{$\mathsf{DARTIC}$ Framework: The architecture integrates identity management (IDML) and crowdsourcing management (CSML) layers with oracle-assisted off-chain data verification to enable privacy-preserving, reputation-driven interactions. The IDML acts as a root identity and credential management ledger that provides Sybil-resistant identity binding and issues contextual credentials used for access control. The CSML represents one or more application-specific (context-driven) ledgers, which may operate under an \textit{open permissioned} policy. Smart contracts (SCs) names are, \textit{ISC: identity, BSC: business, RSC: reputation, ASC: access and DSC: deposit}; \textit{SSA} and \textit{SSN} denote \textit{social security administration} and \textit{number}; \textit{DDP} refers to a \textit{decentralized dispute protocol}. }
		\label{fig:dartic}
	\end{figure}

	\quad This section presents the proposed $\mathsf{DARTIC}$. We first describe the system model and then outline the architecture and interaction workflow (Fig.~\ref{fig:dartic}) that each user follows to ensure privacy preservation and reputation protection. Table~\ref{tab:notations_dartic} summarizes the main notations used throughout the paper.
	
	\subsection{System Model} \label{sec:$DARTC$threats} 
	\noindent \textbf{1. Entities.} $\mathsf{DARTIC}$ comprises four distinct entities:
	\begin{itemize}
		\item \textit{Requesters:} are the initiators of tasks within the system. They define the requirements, parameters, and rewards associated with the tasks.
		\item \textit{Workers:} users who take on tasks posted by requesters. Workers build their reputation within the system based on the quality of their submissions, which influences their ability to access higher-value or more challenging tasks.
		\item \textit{Validators:} a.k.a. \textit{committee nodes} run a blockchain consensus protocol to agree on the validity of blocks containing identity and business transactions.
	\end{itemize}
	
	\quad Requesters and workers may generate multiple pseudonyms; however, all pseudonyms are cryptographically bound to a single long-term master key to mitigate Sybil and reputation attacks. This design enables users to maintain a unique reputation while interacting under different pseudonyms across sessions. Specifically, the IDentity Management Ledger (IDML) is a consortium blockchain that serves as the root identity and credential management layer, providing Sybil-resistant identity binding and issuing contextual credentials for access control. The CrowdSourcing Management Ledger (CSML) comprises one or more application-specific, context-driven ledgers (e.g., crowd-sensing or crowd-learning), which may operate under different permissioning models, including open-permissioned or fully closed settings. In this design, multiple CSML instances can be securely anchored to the same IDML, enabling decentralized and uniform access control across heterogeneous application domains while preserving the operational autonomy of each CSML at the application layer. We further discuss these components in \S\ref{subsec:archi}.

	\color{black}
	\noindent \textbf{2. Adversarial Model.} We assume that all entities, including \textit{requesters}, \textit{workers}, and \textit{validators}, may exhibit faulty or malicious behavior. No participant is assumed to be inherently trusted, and both requesters and workers are considered rational and potentially adversarial, possibly acting individually or colluding to perform coordinated attacks.

	We assume a Byzantine fault-tolerant setting with $n \geq 3f+1$ nodes, which guarantees safety and liveness in the presence of up to $f$ Byzantine faults. In particular, the IDML, CSML, and oracle networks execute the prescribed protocol while tolerating Byzantine behavior under the standard model~\cite{lamport}. Under this assumption, $f < \frac{1}{3}$ participants may deviate arbitrarily from the protocol.
	\color{black}

	\noindent \textbf{3. Security Properties.} Under the above assumptions, we summarize our security properties (analyzed in \S\ref{sec:$DARTC$security}) as follows:
	\begin{itemize} 
		\item \textit{Sybil resistance:} A single user cannot create multiple false identities to exert undue influence or control. 
		\item \textit{Collusion resistance:} The ability to prevent a malicious group of participants from collectively compromising the system. With respect to identity, it ensures that even if a user (prover) and a subset of committee nodes (verifiers) conspire to manipulate the outcome of a decision about the right to access, they cannot succeed without compromising a significant fraction of the system's security assumptions.
		\item \textit{User privacy}: An adversary cannot determine a user's attributes by examining issued credentials, analyzing transaction data during interactions with other users, or observing the ongoing evaluation of interactions.
		\item \textit{Unlinkability:} This property ensures that the identity of a user cannot be linked to any of its pseudonyms, nor can two pseudonyms of the same user be linked to each other. 
		\item \textit{Accountability:} Actions of participants are traceable to their digital identities in a verifiable manner. Any misbehavior can be detected and appropriately sanctioned without compromising honest users’ privacy.  
		\item \textit{Reputation binding:} A user's reputation is tied to its actual behavior. Although users can generate multiple pseudonyms, all are cryptographically linked to a unique access token.
		\item \textit{Forward Reputation binding:} No user should be able to mint/use a reputation token with a reputation score higher than that linked to his/her most recent token.
	\end{itemize}

	\subsection{Architecture} \label{subsec:archi}
	
	We now present $\mathsf{DARTIC}$ design as depicted in Fig.~\ref{fig:dartic} and outline the workflow between users and key components.\\
	
	\noindent \textbf{1. IDentity Management Ledger (IDML).} IDML is a \textit{consortium} blockchain that manages data about Decentralized IDentifiers (DIDs) and acts as \textit{credentials issuer}. DIDs are publicly identifiable endpoints, such as documents, wallets, smart contracts, etc. We adapt the CanDID \cite{candid} approach to uniquely identify legitimate users. The system relies on a PKI-like infrastructure \cite{pki} to support the use of DIDs. Each user manages a master public/private key pair ($M_{pk}, M_{sk}$). The PKI infrastructure then stores the correspondence between the DIDs and the public key. We use a permissioned model to issue credentials, in which a set of selected nodes acts as a committee. Let $CS$ be the committee set with $n$ nodes: $\{cn_i\}^n_{i=1}$. The nodes in the committee jointly store a secret key $sk_{cs}$, which is used to issue credentials. The corresponding public key $pk_{cs}$ is used to verify credentials. Any party (\eg, smart contract, validator) can act as a \textit{credential verifier}. 
	
	\quad IDML derives Sybil-resistant, privacy-preserving decentralized credentials from commonly used legacy data (Step~1 in Fig.~\ref{fig:dartic}). To this end, it first transforms a set of claims referred to as ``pre-credentials'' to a master credential $M_{cred}=(M_{pk}, \sigma_{cred})$     with a privacy-preserving deduplication protocol \cite{candid}. Master credentials are Sybil-resistant because users can only get one credential and are not intended for interactions. Rather, IDML allows users with a valid $M_{cred}$ to create context-based credentials by linking application-specific attributes (attested to by pre-credentials) to $M_{cred}$. The next step, therefore, consists of the creation of these contextual credentials, where each application (\eg, crowdsensing) defines a distinct context $ctx$. For instance, if Alice wants to obtain a credential within $ctx$, she must submit her $M_{cred}$ to the committee, along with a set of required claims specified by $ctx$ (\eg, age over 18). The committee proceeds to validate the claims and issue a credential for $ctx$ through a process similar to that used to issue $M_{cred}$. The detailed registration with credential issuance is formally presented in \S\ref{subSec:access}.
	
	\noindent \textbf{2. Crowdsourcing Management Ledger (CSML).} CSML is an \textit{open permissioned}~\cite{miller2016} blockchain that implements the crowdsourcing logic. Access to CSML ( Step~2-3 in Fig.~\ref{fig:dartic}) is done using contextual (application-based) credentials provided by the IDML committee after successful registration. Therefore, within our design, multiple CSMLs with various business logic can be incorporated with the IDML. For instance, we can have a CSML for sensing and another for learning. Users on each CSML can generate as many pseudonyms as they want (ideally, a new pseudo for each new interaction) to protect their identity. To achieve this while ensuring effective reputation and preventing reputation attacks, users must provide verifiable proofs (see \S\ref{subsec:anonyrep} for the formal constructions).  
	
	\quad In addition to its business logic, CSML also incorporates the reputation model of $\mathsf{DARTIC}$, which evaluates interactions between users and updates their reputation scores. This model must preserve user privacy as reputation or recommendation systems that rely on users’ historical interactions can inadvertently reveal sensitive information. Consequently, only parameters that do not compromise privacy can be used to evaluate interactions. In $\mathsf{DARTIC}$, the sole parameter reflecting past interactions is the user’s current reputation score.
	
	\noindent \textbf{3. Decentralized Oracle Network (DON).} Traditional blockchains are known for \textit{``The Oracle Problem''}, which refers to the problem of securely integrating real-world data on-chain since blockchains primarily deal with information that is native to the network~\cite{don}. Specifically, smart contracts that run ``on-chain'' cannot process external ``off-chain'' data and events to provide the user with functionality that needs to be realized outside the blockchain. Consequently, solving this problem is crucial to guarantee the security of the off-chain data we want to process on-chain (Steps 1.4, 2.3, 2.4 in Fig.~\ref{fig:dartic}). 
	
	\quad Over the past years, various decentralized oracle protocols have emerged to address this limitation. Among them, two protocols, DECO \cite{deco} and Town Crier \cite{towncrier}, have been developed specifically to enable oracle nodes to securely fetch data from off-chain systems while safeguarding user privacy and data confidentiality. DECO uses Multi-Party Computation (MPC) to achieve its integrity and confidentiality properties. Town Crier, on the other hand, relies on a Trusted Execution Environment (TEE) for confidential computing. 
	
	We chose DECO over Town Crier in $\mathsf{DARTIC}$ due to its trustless design. Specifically, DECO enables a single prover to extract verifiable statements from a private TLS session with a web server and make them publicly verifiable by all nodes (as detailed in \S\ref{subSec:access}). Consequently, the entire network can verify the authenticity of the attested data and trigger on-chain execution.

	\section{Decentralized Anonymous Crowdsourcing} \label{sec:$DARTC$math}
	
	\quad $\mathsf{DARTIC}$ uses a set of cryptographic building blocks on top of IDML and CSML to guarantee anonymous reputation and payments. In the following, we first present these cryptographic primitives; then, the proposed protocols for registration, access, anonymous reputation and payments.
	
	\subsection{Cryptographic Building Blocks} \label{sec:bblocks} 
	\quad The main cryptographic building blocks on which the $\mathsf{DARTIC}$ system is built are the following:
	
	\noindent \textbf{Hash Commitment.} A \textit{commitment} scheme is a cryptographic protocol that allows a party, referred to as the committer, to commit to a chosen value without revealing it, while still being able to prove its validity later on \cite{trapcom}.  A \textit{hash commitment} scheme involves the use of cryptographic hash functions to achieve this goal. Let $x$ be the secret value to commit to, $r$ a trapdoor (randomness or key), and $\mathsf{COMM}$ the commitment function. Committing to $x$ involves computing: \(cm = \mathsf{COMM}_r(x) = H(x \| r),\)
	where $H$ is a cryptographic hash function. Hash commitments are designed to fulfill two security properties: \begin{inparaenum}[(i)] 
		\item \textit{Hiding:} Given $cm$, it should be computationally infeasible to determine the original value $x$,
		\item \textit{Binding:} It should be computationally infeasible to find two distinct values $x_1$ and $x_2$ s.t. $\mathsf{COMM}(x_1) = \mathsf{COMM}(x_2)$.
	\end{inparaenum} 
	
	\vspace{0.2 cm}

	\noindent \textbf{Threshold Signature.} A \textit{threshold signature (TS) }~\cite{gennaro2018fast, gennaro2020one} is a distributed multi-party signature protocol in which the signing key is split into $n$ shares, with each share held by a different individual or entity. When the threshold is set at $t$, no group of fewer than $t + 1$ shares can generate a valid signature on behalf of the group. The protocol for a $(t+1, n)$ TS scheme involves the following steps:
	
	\begin{itemize}
		\item $\mathsf{KeyGen}(1^\lambda) \rightarrow (sk, vk, sk_i, vk_i)$: The system is initialized by generating a secret key $sk$ and a verification key $vk$. These keys are distributed among $n$ signing members, each receiving a secret key share $sk_i$ and a corresponding verification key share $vk_i$. This process can be executed without relying on a trusted third party using a Distributed Key Generation (DKG) method \cite{pdkg}.
		
		\item $\mathsf{PartSign}(m, sk_i) \rightarrow \sigma_i$: Given the secret key share $sk_i$ and a message $m$ to be signed, each member computes a partial signature share $\sigma_i$.
		
		\item $\mathsf{VerifyPartSign}(m, vk_i, \sigma_i) \rightarrow 0/1$: This step verifies the validity of a partial signature. Given the message $m$, the verification key share $vk_i$, and the partial signature $\sigma_i$, the function returns 1 if $\sigma_i$ is valid; otherwise, it returns 0.
		
		\item $\mathsf{AggSign}(\sigma_1, \dots, \sigma_{t+1}) \rightarrow \sigma$: Once $t+1$ valid partial signatures $\sigma_1, \dots, \sigma_{t+1}$ are collected, they are combined to produce the final aggregated signature $\sigma$.
		
		\item $\mathsf{VerifySign}(vk, m, \sigma) \rightarrow 0/1$: In the final step, the validity of the aggregated signature $\sigma$ is checked. Given the verification key $vk$, the message $m$, and the aggregated signature $\sigma$, the function outputs 1 if  $\sigma$ is valid; otherwise, it outputs 0.
	\end{itemize}
	
	A threshold signature scheme satisfies the following:
	\begin{inparaenum}[(i)]
		\item \textit{Unforgeability}: Given any $t$ partial signatures, an adversary is unable to forge a valid, complete signature for a message that has not been signed by the honest participants.
		\item \textit{Robustness}: If the adversary controls $\le t$ members, the honest participants can still generate the complete signature.
	\end{inparaenum}
	
	\noindent \textbf{zkSNARKs.} A \textit{zero-knowledge Succinct Non-Interactive Argument of Knowledge (zkSNARKs)} is a Non-Interactive Zero-Knowledge (NIZK) scheme \cite{groth}, wherein the proof itself is a self-contained data block that can be verified without requiring any interaction from the prover. It provides zero-knowledge, soundness, completeness, succinctness, and non-interactivity as key properties \cite{plonk}. 
	
	\quad A zkSNARK construction \cite{groth} consists of four algorithms ($\mathsf{Setup}$, $\mathsf{Gen}$, $\mathsf{Prov}$, $\mathsf{Verif}$) defined as follows:
	\begin{itemize}
		\item The $\mathsf{Setup}$ algorithm takes a security parameter $\lambda$ and generates the public parameters: $\mathsf{Setup}(1^\lambda)\rightarrow pp  $
		
		\item The key generator algorithm $\mathsf{Gen}$ takes $pp$ and a program $C$, and generates two publicly available keys: a proving key $PK$, and a verification key $VK$; $(PK,VK) = \mathsf{Gen}(pp, C)$. These keys are public parameters that need to be generated only once for a given program $C$.
		\item The proving algorithm $\mathsf{Prov}$ takes as input the proving key $PK$, a public input $t$, and a private witness $w$. The algorithm generates a proof $\pi = \mathsf{Prov}(PK, t, w)$ that the prover knows a witness $w$ and that the witness satisfies the program $C$.
		\item The verification algorithm computes $\mathsf{Verif}(VK, t, \pi)$ which returns true if the proof is correct, and false otherwise. Thus, this function returns true if the prover knows a witness $w$ satisfying $C$.
	\end{itemize}

	\noindent \textbf{zkSet Membership via Merkle Trees.} This involves proving that an element belongs to a set using the Merkle tree data structure. Formally, given a set $S$ containing $n$ elements and a Merkle tree constructed from the hash values of these elements, the set membership problem is to prove that a specific element $x$ belongs to the set $S$ without revealing any other elements in $S$ \cite{zkset}. Merkle trees alone do not provide zk property. To achieve this, we combine Merkle trees with zkSNARK constructions\cite{groth,plonk}, ensuring:  
	\begin{inparaenum}[(i)]
		\item \textit{Correctness}: The proof is valid if and only if $x$ truly belongs to $S$, 
		\item \textit{Zero-Knowledge}: The verifier learns nothing about $S$ or $x$ beyond the validity of the proof,  
		\item \textit{Succinctness}: The proof size and verification time are constant $O(1)$.  
	\end{inparaenum}
	
	\quad We define two variants of the zkSet problem to ensure both privacy-preserving access control and reputation binding.
	\begin{definition}[zkSet for Access Control] \label{def1}
		A prover proves knowledge of a secret \( a \) s.t. the cryptographic commitment \( cm_A \), representing the user's access credentials $A$, appears as a leaf in a collision-resistant hash (CRH)-based Merkle tree, \( ACTree \), whose root is \( rt_A \). 
	\end{definition}
	
	\begin{definition}[zkSet for Reputation Binding]\label{def2}  
		A prover proves knowledge of a secret \( b \) s.t. the cryptographic commitment \( cm_R \), computed based on a reputation score \( R \), appears as a leaf in a CRH-based Merkle tree, \( RCTree \), whose root is \( rt_R \). 
	\end{definition}
	
	We detail the integration of these variants in \S\ref{subsec:anonyrep} and how they are scaled through aggregation and batching in \S\ref{sub:scalability}.

	\subsection{Registration \& Access} \label{subSec:access}
	\quad The registration process takes place between a user $u$ and the IDML committee. We build a trustless protocol that ensures uniqueness while issuing master credentials $M_{cred}$ for any valid user $u$. To this end, we use a decentralized Oracle network to enable the import of credentials from legacy systems. For example, Alice can use her credentials on her Social Security Administration (SSA) account to generate a pre-credential certifying her Social Security Number (SSN) \cite{candid}. We extend the DECO protocol~\cite{deco} with verifiable randomness and zero-knowledge proofs (ZKPs) to preserve user privacy while ensuring decentralization. 
	
	\quad DECO involves a prover~(\(P\)), a verifier~(\(V\)), and a TLS channel, enabling \(P\) to convince \(V\) that private server data satisfies a given predicate. We enhance DECO’s multiparty computation (MPC) with verifiable randomness to guarantee authenticity and integrity, and integrate zero-knowledge proofs (ZKPs) to attest predicate satisfaction without revealing the underlying data. The process involves running a multiparty computation (MPC) protocol, with a user $u$ acting as the \textit{prover} and attesting to the $claim$: ``the webpage retrieved from the SSA website includes a private string \eg, SSN: 123-45-6789''. Protocol~\ref{protocol:register} details the required steps. 
	
	\refstepcounter{protocol} 
	\begin{protocolbox}{$\mathcal{U} \xrightarrow[\text{}]{\texttt{Register}}$ $\mathbf{IDML}$ \{$ISC$\}, $\mathbf{DON}$}
		\label{protocol:register}
		\begin{small} 
			\begin{itemize}
				\item[\scriptsize{$\blacksquare$}] \textbf{Setup:} IDML \( CS = \{ cn_i \} \)  with signing keys \( \{ sk_i \} \).
				\item[\scriptsize{$\blacksquare$}] \textbf{Result:} Issue valid master credentials $M_{cred}$ for user $u$. 
			\end{itemize}
			\begin{enumerate} 
				\item Initialization: User \( u \) invokes the contract $ISC$ to register.
				
				\item Random Committee Selection: 
				\begin{itemize}
					\item  \( DON \) receives an on-chain event $\texttt{REGISTER}$.
					\item Then generates a random seed \( r \) using a Verifiable Random Function (VRF)\cite{vrf}, producing \( (r, \pi_{r}) \). 
					\item \( \pi_{r} \) is broadcasted and verified by DON and IDML nodes to ensure \( r \) is authentic.
					
					\item The random seed \(r\) is used to select \(k > t+1\) committee nodes \(\{cn_1, \dots, cn_k\}\): 
					\(cn_i = H^i(r) \bmod n, \quad i = 1, \dots, k\). Here, \(H^i(r)\) denotes applying \(H\) iteratively \(k\) times.
					
				\end{itemize}
				
				\item Proof Verification and Partial Signing: 
				\begin{itemize}
					\item For each \( cn_i \) (\( i \in [1, k] \)): 
					\begin{itemize}
						\item \( u \) executes Algo.\ref{claimdeco} to prove the $claim$ $\Phi$ with \( cn_i \) as the verifier.
						\item \( cn_i \) verifies the received proof $\Pi$  
						\item If valid, \( cn_i \) generates a partial ECDSA signature \( \sigma_{i} = \text{Sig}_{sk_i}(\Phi) \) and sends it to \( DON \)
					\end{itemize}
				\end{itemize}
				
				\item Signature Aggregation: 
				\begin{itemize}
					\item \( DON \) aggregates the partial signatures \( \{ \sigma_{i} \} \) to produce the final committee signature \( \sigma_{cred}\). 
				\end{itemize}
				
				\item On-Chain Verification: 
				\begin{itemize}
					\item \( DON \) invokes the \texttt{verifyClaim} function within $ISC$ using \( \sigma_{cred} \) to validate the claim publicly on-chain
				\end{itemize}
				
			\end{enumerate}
		\end{small}
	\end{protocolbox}
	
	\begin{algorithm}[t]
		\footnotesize
		\DontPrintSemicolon
		
		\caption{Proving Claim via DECO~\cite{deco} and ZKPs} \label{claimdeco}
		\KwData{Website WS, $claim$ $\Phi$, Verifier $cn_i$}
		\KwResult{Zero-Knowledge Proof $\Pi$ of $claim$}
		\Begin{
			Prover $u$ and verifier $cn_i$ execute the DECO \textit{three-party handshake} with WS, establishing secret-shared TLS session keys $(k_P, k_V)$\;
			$u$ retrieves encrypted data $D$ from WS via the DECO \textit{query execution protocol}\;
			$u$ commits to the TLS transcript and obtains verifier key share\;
			$u$ computes commitment $\mathsf{COMM}(D)$\;
			$u$ formulates the claim $\Phi(D) =$ "SSN matches criteria"\;
			$u$ generates ZKP $\Pi$ proving that:
			\begin{itemize}
				\item $D$ originates from an authentic TLS session
				\item $\mathsf{COMM}(D)$ corresponds to the committed transcript
				\item $\Phi(D)$ holds
			\end{itemize}
			$u$ sends $(\Phi, \Pi)$ to verifier $cn_i$
			
		}
		
	\end{algorithm}
	If the \( \sigma_{cred}\)  is valid, the master public key $M_{pk^u}$ of the user is whitelisted. $u$ then can post another set of claims to IDML to get access to CSML using context-based credentials~\cite{candid}. To obtain a new credential for the $ctx$ context (\eg, ``crowdsensing''), $u$ must submit to the committee ($pk_{ctx}, M_{cred}, \Phi^u_{ctx}$), a new identifier $pk_{ctx}$ to be used in the $ctx$ context,  master credentials and a set of pre-credentials with the claims $\Phi^u_{ctx}$ required by $ctx$ (\eg, location = ``Paris city''). The IDML via the contract $ISC$ upholds a set of $Granted_{ctx}$ identifiers denoting those that have already obtained a credential within this specific context. If $M_{pk^u}$ of $M_{cred}$ is not part of this set, a credential is issued. Finally, ($M_{pk^u} , pk^u_{ctx}$) is added to $Granted_{ctx}$. To manage access tokens efficiently, we use a CRH-based Merkle tree (with root $rt_A$) called $ACTree$. The tree stores all the users' access commitments as leaves. hence, minting a new access token \texttt{AT} means adding a new valid leaf $cm_A$ to $ACTree$. This is done following Protocol~\ref{protocol:mintat}.
	
	\refstepcounter{protocol} 
	\begin{protocolbox}{$\mathcal{U} \xrightarrow[\text{}]{\texttt{MintAT}}$ $\mathbf{IDML}$ \{$ASC$\}}
		\label{protocol:mintat}
		\begin{small} 
			\begin{itemize}
				\item[\scriptsize{$\blacksquare$}] \textbf{Data:} $ACTree$ with root $rt_A$, contains all user access commitments. 
				\item[\scriptsize{$\blacksquare$}] \textbf{Result:} Generate and Mint Access Token. 
			\end{itemize}
			\begin{enumerate} 
				
				\item  User $u$ generate key pair $(pk, sk)$ public and secret key.
				\item  Compute access commitment in two steps: 
				\begin{enumerate}[leftmargin=1em, label=\roman*)]
					\item Sample a random $r$ and set: $cm_u = \mathsf{COMM}_r(sk^u_{ctx})$
					\item Sample $r'$ and set:
					$  cm_A = \mathsf{COMM}_{r'}(cm_u || pk)$

				\end{enumerate}
				
				\item  Define access token:
				$\texttt{AT} = (sk^u_{ctx}, r, r', cm_A)$
				\item  Create transaction: $\texttt{TX\textsubscript{AM}} = (pk, cm_u, cm_A)$

				\item Invoke $ISC$: submit \texttt{TX\textsubscript{AM}} to the IDML \\
				- Only accept if $pk^u_{ctx}$ is known
				to the committee
				\item  Update $ACTree$ with new $cm_A$ and emit $\texttt{MINTAT}$ event
				\item \textit{Synchronization:} Upon receipt of $\texttt{MINTAT}$ event, DON updates $ACTree$ on CSML with new leaf $cm_A$
			\end{enumerate}
		\end{small}
	\end{protocolbox}

	The protocol ensures that master credentials are never used in interactions, preventing any linkage to the user's real identity. Furthermore, contextual credentials are used exclusively within the IDML and remain cryptographically concealed when handled by the CSML.
	
	\begin{figure}
		\centering
		\includegraphics[width=0.95\linewidth]{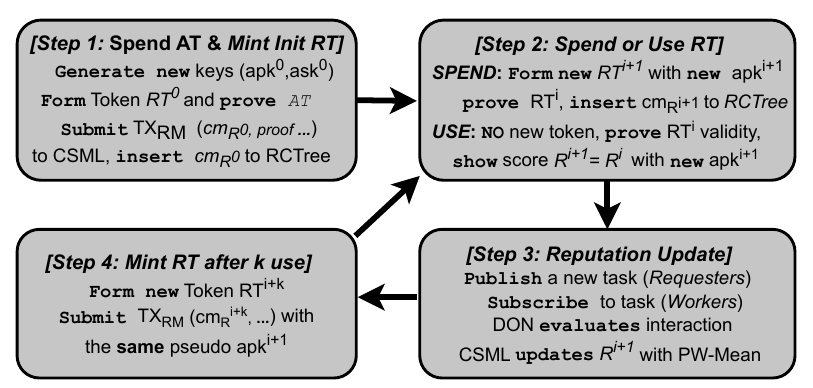}
		\caption{$\mathsf{DARTIC}$'s anonymous reputation lifecycle. Users start by \textbf{(Step 1)} proving the ownership of a valid $\texttt{AT}$ and minting an initial $\texttt{RT}^0$. In \textbf{(Step 2)}, reputation tokens are spent or used to interact within the CSML, generating unlinkable pseudonyms for each transaction. In \textbf{(Step 3)}, oracle nodes evaluate interactions and securely update reputations on-chain via a threshold signature. Finally, in \textbf{(Step 4)}, after $k\geq 1$ updates, users can mint new tokens ($\texttt{RT}^{i+k}$) to restore anonymity while carrying forward accumulated reputation.}
		
		\label{fig:anonyrep}
	\end{figure}
	\vspace{-0.2cm}

	\subsection{Anonymous Reputation} \label{subsec:anonyrep}
	\quad Now that we have described the registration process, we will move on to anonymous reputation, which consists of four key steps as depicted in Fig.~\ref{fig:anonyrep}. 
	
	\subsubsection*{\textbf{Step 1. Mint Initial Reputation Token}}
	In this initial step, a reputation token is minted that is cryptographically linked to the user's contextual credentials. This process conceals the user's digital identity (\ie, $pk^u_{ctx}$) while granting access to the CSML. To do so, users must spend their access tokens and mint their initial reputation token $\texttt{RT\textsuperscript{0}}$ following Protocol~\ref{protocol:spendat}. This results in a new valid leaf being inserted into $RCTree$, containing a commitment to the initial reputation score $R^0$.
	
\begin{figure}[t]
		\centering
		\resizebox{\columnwidth}{!}{
			\begin{tikzpicture}[
				every node/.style={font=\Large},
				box/.style={draw, rounded corners, align=center},
				leaf/.style={draw, minimum width=2.2cm, minimum height=0.7cm},
				>=latex
				]
				\draw[decorate,decoration={brace,amplitude=6pt},thick]
				(2.8,0) -- (2.8,4);
				
				\node[left] at (2.6,2)
				{$L$};
				
				\node[box] (rt) at (0,4)
				{\texttt{RT}$^i$};
				
				\node[box] (cmp) at (0,2.5)
				{$cm_p^i$\\
					$\mathsf{COMM}(apk^i \Vert s^i)$};
				
				\node[box] (cmr) at (0,0.4)
				{$cm_R^i$\\
					$\mathsf{COMM}(R^i \Vert cm_p^i)$};
				
				\draw[->] (rt) -- (cmp);
				\draw[->] (cmp) -- (cmr);
				
				\node[leaf] (c0) at (4,0) {$cm_R^0$};
				\node[leaf] (c1) at (6.5,0) {$cm_R^1$};
				
				\node[leaf] (c2) at (9,0) {$cm_R^i$};
				\node[leaf] (c3) at (11.5,0) {$cm_R^{i+1}$};
				
				\node[circle,draw] (h1) at (5.5,2) {$H$};
				\node[circle,draw] (h2) at (10,2) {$H$};
				
				\node[circle,draw] (root) at (7.75,4) {$H$};
				
				\draw[->] (c0) -- (h1);
				\draw[->] (c1) -- (h1);
				
				\draw[->] (c2) -- (h2);
				\draw[->] (c3) -- (h2);
				
				\draw[->,dashed] (h1) -- (root);
				\draw[->, dashed] (h2) -- (root);

				\node[above=0.5cm] at (root)
				{$rt_R$};
				
				\node[box] (zk) at (13,4)
				{zkSet Proof};
				
				
				\draw[dashed,->] (root) -- (zk);
				
				\node[box] (spend) at (13,1.5)
				{SpendRT\\UseRT};
				
				
				\draw[->] (zk) -- (spend);
				
				\draw[red,thick,->, solid] (c2) -- (h2);
				\draw[red,thick,->, dashed] (h2) -- (root);
				
				\node[red,right] at (8.7,3.6)
				{$\pi_R^i$};

		\end{tikzpicture}}
		
		\caption{
			\texttt{RT} structure and its integration into the commitment tree. An \texttt{RT} contains a nested commitment $cm_R=\mathsf{COMM}(R\Vert cm_p)$, where $cm_p=\mathsf{COMM}(apk\Vert s)$ binds the pseudonym and serial number. Each minted \texttt{RT}$^i$ contributes a new leaf $cm_R^i$ to the \textit{RCTree} of depth $L$. To spend or use a token, the owner generates a zkSet proof $\pi_R^i$ (Def.~\ref{def2}) demonstrating knowledge of a valid membership path/witness, corresponding to a leaf $cm_R^i$, that verifies against the tree root $rt_R$. This proves that the committed \texttt{RT}$^i$ is included in the \textit{RCTree} without revealing its position or linking it to its owner.}
		\label{fig:rctree}
	\end{figure}
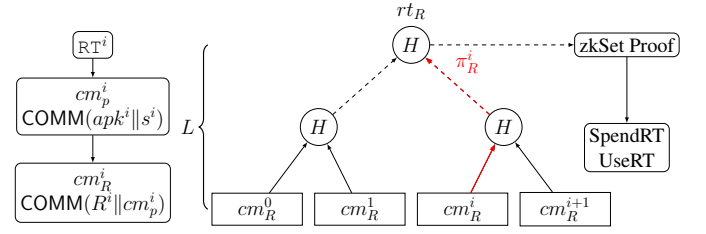

	\quad To provide targets for new tokens, we use addresses: each user $u$ generates an address key pair ($apk, ask$), the address public and private key, respectively. The token of $u$ contains the value $apk$ and can only be used with the knowledge of $ask$. A key pair ($apk, ask$) is sampled by choosing a random seed $ask$ and setting $apk := \mathsf{PRF}_{ask} (0)$ using a Pseudo-Random Function. To ensure forward reputation binding, the user must also sample a random serial number, $S_n$, for each new reputation token as shown in Fig.~\ref{fig:rctree}. This serial number is then revealed when the token is used. 
	
	The anonymity of the user is achieved because the proof $\pi_{AS}$ is zero-knowledge: while $cm_{u}$ and $pk$ are revealed, no information about $r'$ is revealed, and finding which of the many commitments in $ACTree$ corresponds to \texttt{TX\textsubscript{AS}} is equivalent to inverting $f(r'):= \mathsf{COMM}_{r'}(X)$, which is computationally infeasible \cite{trapcom}.
	
	\refstepcounter{protocol} 
	\begin{protocolbox}{$\mathcal{U} \xrightarrow[\texttt{}]{\texttt{SpendAT}}$ $\mathbf{CSML}$ \{$ASC$, $RSC$\}}
		\label{protocol:spendat}
		\begin{small}
			\begin{itemize}
				\item[\scriptsize{$\blacksquare$}] \textbf{Data:} $\texttt{AT}$, $ACTree$ with root $rt_A$, initial reputation $R^0$
				\item[\scriptsize{$\blacksquare$}] \textbf{Result:} Spend \texttt{AT} and Mint Initial \texttt{RT\textsuperscript{0}}. 
			\end{itemize}
			
			\begin{enumerate}
				\item User \( u \) generates a new key pair \((apk, ask)^0\)
				\item User samples a random \( s^0 \) and set: $S_n^0 := [\mathsf{PRF}_{ask}(s)]^0$
				
				\item User samples two random $r_1^0$ and $r_2^0$ and commits to the tuple \((apk, R, s)^0\) in two steps: i) $cm_p^0 := [\mathsf{COMM}_{r_1}(apk || s)]^0$, ii) $cm_R^0 := [\mathsf{COMM}_{r_2}(R || cm_p)]^0$
				\item This results in: $\texttt{RT\textsuperscript{0}} := (apk, R, s, r_1, r_2, cm_R)^0$
				
				\item User $u$ produces a zk proof \( \pi_{AS} \) for the NP statement (Def.~\ref{def1}) :
				\begin{quote}
					\textbf{\textit{``I know a secret \( r' \) s.t. \( \mathsf{COMM}_{r'}(cm_{u} || pk) \) appears as a leaf in a CRH-based Merkle Tree \( ACTree \) whose root is \( rt_A \)''}}
				\end{quote}
				\item Then defines: $\texttt{TX\textsubscript{AS}} := (R^{0}, cm_p^0, r_2, cm_R^0, \pi_{AS})$ 
				
				\item Invoke $ASC$: submit \( \texttt{TX\textsubscript{AS}} \) to the CSML.\\
				- Accept only if $\pi_{AS}$ is valid and $cm_R^0$ is correct.
				\item Update $RCTree$ with new leaf $cm_R^0$.
			\end{enumerate}
		\end{small}
	\end{protocolbox}

	\subsubsection*{\textbf{Step 2. Reputation Token Spend/Use}}
	So far, user \textit{u} has minted his initial reputation token $\texttt{RT\textsuperscript{0}}$. Therefore, $u$ can interact with any other user \textit{v} on the CSML by submitting transactions. Within $\mathsf{DARTIC}$, users' reputation scores are tied to their most recent reputation commitment $cm_R$. Thus, for interaction, users must reveal this nested commitment to disclose their reputation scores. Since only the user possessing the secret $r'$ can show it, there is no risk of forgery. 

	\textit{Spend A Reputation Token.} To adopt a new pseudonym in future interactions---denoted $apk^{i+1}$, distinct from the current $apk^i$---a user $u$ on CSML spends his reputation token. This is done by submitting an RT spending transaction, $\texttt{TX\textsubscript{RS}}$, as defined in Protocol~\ref{protocol:spendrt}. This transaction enables $u$ to generate a new token $\texttt{RT}^{i+1}$ of identical value ($R^{i+1}=R^i$) to the current one. The user must attach a zk proof $\pi_{RS}$ to prove the validity of both the old and new token, and the fact that a valid commitment $cm_R$ exists as a leaf in $RCTree$.
	
	\refstepcounter{protocol} 
	\begin{protocolbox}{$\mathcal{U} \xrightarrow[\text{}]{\texttt{SpendRT}}$ $\mathbf{CSML}$ \{$RSC$\}}
		\label{protocol:spendrt}
		\begin{small}
			\begin{itemize}
				\item[\scriptsize{$\blacksquare$}] \textbf{Data:}  old key pair $(apk, ask)^{i}$,  old reputation token $\texttt{RT\textsuperscript{i}} := (apk, R, s, r_1, r_2, cm_R)^i$,  and $rt_R$.
				\item[\scriptsize{$\blacksquare$}] \textbf{Result:} Spend old \texttt{RT\textsuperscript{i}} and Mint new \texttt{RT\textsuperscript{i+1}}.
			\end{itemize}
			\textbf{a.} Produce a new token \texttt{RT\textsuperscript{i+1}}:
			\begin{enumerate}
				\item User \( u \) generates a new key pair \((apk, ask)^{i+1}\)
				\item User $u$ samples serial number randomness $s^{i+1}$
				
				\item Set $cm_p^{i+1} := [\mathsf{COMM}_{r_1}(apk||s)]^{i+1}$ with random $r_1^{i+1}$
				\item Set $cm_R^{i+1} := [\mathsf{COMM}_{r_2}(R||cm_p)]^{i+1}$ with random $r_2^{i+1}$
				\item Form $\texttt{RT\textsuperscript{i+1}} := (apk, R, s, r_1, r_2, cm_R)^{i+1}$   
				
			\end{enumerate}
			
			\textbf{b.} Generate zk proof $\pi_{RS}$  to spend old token \texttt{RT\textsuperscript{i}} that:
			\begin{enumerate}
				\item  Verify well-formedness of tokens:
				\begin{itemize}
					
					\item  Check $cm_p^{i} == [\mathsf{COMM}_{r_1}(apk||s)]^i$ and $cm_R^{i} == [\mathsf{COMM}_{r_2}(R||cm_p)]^i$ 
					\item Similarly, check $cm_p^{i+1}$ and $cm_R^{i+1}$
				\end{itemize}
				\item  Verify key matching:
				\begin{itemize}
					\item $apk^{i} == \mathsf{PRF}_{ask^{i}} (0)$ and $apk^{i+1} == \mathsf{PRF}_{ask^{i+1}} (0)$
				\end{itemize}
				\item Verify correct serial number: $S_n^{i} == \mathsf{PRF}_{ask^{i}} (s^{i})$
				\item Check $cm_R^{i}$ is a leaf in $RCTree$ with root $rt_R$ (Def.~\ref{def2})
				\item Check reputation values match: $R^{i+1} == R^{i}$
			\end{enumerate}   
			\textbf{c.} Define $\texttt{TX\textsubscript{RS}}  := (rt_R, S_n^{i}, cm_R^{i+1}, \pi^i_{RS})$ 
			
			\textbf{d.} Invoke $RSC$: submit \texttt{TX\textsubscript{RS}} to the CSML. \\
			- Reject if $S_n^{i}$ appears in a prior transaction or $\pi^i_{RS}$ is invalid.
		\end{small}
	\end{protocolbox}
	
	\textit{Use A Reputation Token.} Protocol~\ref{protocol:usert} details how to use $\texttt{RT\textsuperscript{i}}$: 
	\refstepcounter{protocol} 
	\begin{protocolbox}{$\mathcal{U} \xrightarrow[\text{}]{\texttt{UseRT}}$ $\mathbf{CSML}$ \{$RSC$\}}
		\label{protocol:usert}
		\begin{small}
			\begin{itemize}
				\item[\scriptsize{$\blacksquare$}] \textbf{Data:}  $RCTree$ with root $rt_R$, $S_n^{i}$, $cm_R^{i+1}$, $\texttt{RT\textsuperscript{i}}, ask^{i}$
				\item[\scriptsize{$\blacksquare$}] \textbf{Result:} Use current \texttt{RT\textsuperscript{i}}.
			\end{itemize}
			\begin{enumerate}
				\item User \( u \) generates a new key pair \((apk, ask)^{i+1}\)
				\item Generate zk proof $\pi_{RU}$ to use \texttt{RT\textsuperscript{i}} that:
				\begin{enumerate}[label=\roman*),leftmargin=0.5em,itemsep=2.5pt, parsep=2.5pt]
					
					\item Verify \texttt{RT\textsuperscript{i}}: $cm_p^{i} == [\mathsf{COMM}_{r_1}(apk||s)]^i$ and $cm_R^{i} == [\mathsf{COMM}_{r_2}(R||cm_p)]^i$ 
					\item Check $apk^{i} == \mathsf{PRF}_{ask^{i}} (0)$ and  $S_n^{i} == \mathsf{PRF}_{ask^{i}} (s^{i})$
					\item Check $cm_R^{i}$ is in $RCTree$ with root $rt_R$ (Def.~\ref{def2})
				\end{enumerate}
				\item Create transaction $\texttt{TX\textsubscript{RU}} =(rt_R, S_n^{i}, R^{i}, \pi^i_{RU})$
				\item Invoke $RSC$: submit \texttt{TX\textsubscript{RU}} to CSML.\\
				- Accept only if $\pi^i_{RU}$ is valid and $S_n^{i}$ does not appear in a prior TX. If so, add $(apk^{i+1}, R^{i+1} = R^{i})$ to $Valid$.
				
			\end{enumerate}
		\end{small}
	\end{protocolbox}
	
	This allows $u$ to show the most recent reputation $R^i$ and link it to a new pseudo $apk^{i+1}$ (without minting a new token) by proving $\texttt{RT\textsuperscript{i}}$ correctness and a valid $cm_R^{i}$ is in $RCTree$. A successful \texttt{TX\textsubscript{RU}} allows $u$ to interact with other users. The corresponding $apk^{i+1}$ is then added to 
	$Valid$, a whitelist that maintains the set of authorized application public keys together with their associated reputation values. The user’s identity is protected, as $pk^{u}_{ctx}$ remains hidden. The access token spending transaction, \texttt{TX\textsubscript{AS}}, is submitted under a pseudonym $apk^i$ $\ne$ $pk^{u}_{ctx}$. Likewise, \texttt{TX\textsubscript{RS}} and \texttt{TX\textsubscript{RU}} transactions are posted with a new pseudonym $apk^{i+1}\ne apk^i$. The three pseudonyms are mutually unlinkable as both transactions reveal nothing but a random $S^i_n$, the Merkle root $rt_R$, and a zk proof. The completeness, soundness and zk properties are formally established in Appendix~B.

	\subsubsection*{\textbf{Step 3. Reputation Update}} This step is automated through DON integration. First, the network retrieves the relevant off-chain data and derives the evaluation metrics (via an oracle function). After collective verification and attestation, the DON runs Algo.~\ref{algReputation} which invokes the RSC to update the reputations of $u$ and $v$ under their pseudonyms $apk_i^{u}$ and $apk_i^{v}$. The signature here is instantiated using a standard $(t+1,n)$ ECDSA scheme~\cite{gennaro2018fast, gennaro2020one} as defined in~\ref{sec:bblocks}, to ensure both interoperability and strong threshold security. Concretely, the oracle nodes execute a DKG~\cite{pdkg} protocol to jointly generate shares of a single ECDSA private key without relying on any trusted third party. During reputation updates, each oracle produces a partial signature on the task identifier and associated metrics, which are then verified and securely aggregated into a final signature $\sigma_{orc}$ once at least $t+1$ valid shares are collected. Importantly, the resulting aggregated signature is indistinguishable from a standard ECDSA signature and can be verified using the corresponding public key by the smart contract. This makes the protocol fully compatible with well-established constructions (\eg, GG18~\cite{gennaro2018fast}, GG20~\cite{gennaro2020one}), which provide security against up to $t$ colluding adversaries. The complete reputation update process is detailed next.

	\paragraph*{Interaction Participation} Protocol~\ref{protocol:subtask} outlines how a requester $u$ and a worker $v$ share and participate in a task.
	
	\refstepcounter{protocol} 
	\begin{protocolbox}{$\mathcal{U} \xrightarrow[\texttt{SubscribeTask}]{\texttt{PublishTask}}$ $\mathbf{CSML}$ \{$BSC$\}}
		\label{protocol:subtask}
		\begin{small}
			\begin{itemize}
				\item[\scriptsize{$\blacksquare$}] \textbf{Data:}  $taskID$, $reward$, $sk_{ctx}^u$ for $u$ and $sk_{ctx}^v$ for $v$
				\item[\scriptsize{$\blacksquare$}] \textbf{Result:} Publish/Subscribe to a Task
			\end{itemize}
			\begin{enumerate}
				\item The user $u$ computes $H^u := \mathsf{COMM}(taskID || sk_{ctx}^u)$
				
				\item The user $u$ generates a zkSNARK proof $\pi_{I}^u$ for the following NP statement:
				\begin{itemize}
					\item[] \textbf{\textit{``Given the task identifier $taskID$, I know a secret $sk_{ctx}^u$, and \texttt{AT}$^u$ s.t. $H^u$ is computed correctly and shares the same $sk_{ctx}^u$ with \texttt{AT}$^u$''}}.
				\end{itemize}
				
				\item Invoke $BSC$: $\texttt{TX\textsubscript{newTask}} := (taskID, reward, H^{u}, \pi_{I}^u)$.
				\item Similar to $u$, user $v$ runs steps 1 and 2. Then, 
				$v$ submits $\texttt{TX\textsubscript{subToTask}} := (taskID, H^{v}, \pi_{I}^v)$ to the CSML.
				
				\item $\texttt{TX\textsubscript{newTask}}$ is rejected if $\pi_I^u$ is invalid; and  $\texttt{TX\textsubscript{subToTask}}$ is rejected if $H^v == H^u$ or the proof $\pi_I^v$ is invalid.
				
			\end{enumerate}
		\end{small}
	\end{protocolbox}

	Suppose an interaction where a worker $v$ wants to perform a task (\ie, provide a service) for a requester $u$. The requester $u$ must first publish the task. To do so, $u$ must use the current reputation token (\textbf{Step 2}). Then $u$ can share the new task by posting the transaction $\texttt{TX\textsubscript{newTask}} := (taskID, reward, ...)$,  $taskID$ can be the content identifier (CID) of the task description on IPFS~\cite{ipfs}. Now, any worker $v$ who wants to perform this task must subscribe to the task by posting a \texttt{TX\textsubscript{subToTask}} transaction. Both $u$ and $v$ must provide zkSNARK proof that they do not share the same access token. This condition is necessary for our construction because it prevents self-promotion attacks.

	\begin{algorithm}[t]
		\footnotesize
		\DontPrintSemicolon
		\caption{Distributed Reputation Token Update.\label{algReputation}}
		\KwData{$taskID$, Threshold $t$, $RSC$ address, $requester$, $workers$, metrics}
		\KwResult{Reputation tokens updated for requester and workers.}
		
		\SetKwProg{Upon}{Upon}{ do}{end}
		\SetKwFunction{Fverify}{verifySignature}
		\SetKwFunction{Fupdate}{updateReputation}
		\SetKwFunction{FexecuteUpdateRT}{executeUpdateRT}
		\SetKwProg{Fn}{Function}{:}{}
		\SetKwProg{Pn}{Function}{:}{}
		
		\BlankLine
		
		\Comment{Any oracle node} \;
		\Upon{receipt of \textsc{taskStatusEvent}($taskID$)}{
			\If{\textsc{getStatus}($taskID$) == "Valid"}{
				Broadcast \textsc{updateRTRequest}($taskID$, $requester$, $workers$, metrics) to all oracles\;
			}
		}
		
		\Comment{The aggregator is the node that distributed the request}\;
		\Upon{receipt of \textsc{updateRTRequest}($taskID$, $requester$, $workers$, metrics)}{
			\If{\textsc{validateRequest}($taskID$)}{
				$\sigma_i \gets \textsc{generatePartialSignature}(taskID)$\;
				Send \textsc{partialSignature}($\sigma_i$) to aggregator\;
			} \Else {
				Send \textsc{validationFailed} to aggregator\;
			}
		}
		
		\Upon{receipt of \textsc{partialSignature}($\sigma_i$) from oracles}{
			$S \gets S \cup \{\sigma_i\}$\;
			\If{$|S| \geq t +1 $}{
				$ \sigma_{orc} \gets \textsc{combineSignatures}(S)$\;
				Broadcast ($\sigma_{orc}$, $taskID$) to oracles\; 
				Call $RSC\textsc{.updateRT}(requester, workers, \text{metrics})$\;
			}
		}
		
		\Comment{On-chain verification and RT update}\;
		\Pn{\FexecuteUpdateRT{$\sigma_{orc}, taskID, requester, workers, \text{metrics}$}}{
			\lIf{\textsc{verifyWithPublicKey}($\sigma, taskID$)}{
				$ \textsc{updateRT}(requester, workers, \text{metrics})$}
		}
	\end{algorithm}

	\paragraph*{Interaction Evaluation} The evaluation of interactions must strictly preserve privacy: the reputation model neither relies on nor exposes workers’ historical or sensitive data. Each interaction is assigned a weight defined as:  
	\vspace{-0.3pt}
	\begin{equation} \label{equ1}
		\left\{ \begin{array}{c}
			\mathsf{W_f} = \kappa \cdot \big[ \omega_0 + \omega_d F_d + \omega_a F_a + \omega_{st} C_{st} \big] \\[6pt]
			\kappa, \omega_0,\omega_d,\omega_a,\omega_{st} \in [0,1], \\[4pt]
			\omega_0 + \omega_d + \omega_a + \omega_{st} = 1,
		\end{array}\right.
	\end{equation}
	\vspace{-0.2pt}
	where $\kappa$ is a scaling factor, $\omega_0$ is the baseline weight, and $\omega_d$ and $\omega_a$ capture the normalized duration $d$ and amount $a$, respectively: $F_d = \min(\tfrac{d}{d_{\max}}, 1), 
	\quad
	F_a = \min\!(\tfrac{a}{a_{\max}}, 1)$
	
	The optional term $\omega_{st} C_{st}$ accounts for spatiotemporal context. By design, all contributions are capped within $[0,1]$, to mitigate collusion and Sybil-style inflation attacks. Here, $d_{\max}$ denotes the maximum meaningful interaction duration (\eg, $24$ hours or a domain-specific bound), and $a_{\max}$ is the maximum meaningful amount (\eg, the highest transaction value). Interactions with negligible duration or value (\eg, $d = 0.1$ h, $a = 1$ unit) yield near-zero $F_d$ and $F_a$, ensuring that repeated micro-interactions cannot be exploited.  
	
	The worker $v$ contribution is evaluated by combining rater feedback with auxiliary contextual metrics:  
	\begin{equation} 
		T^{v} = \sum \alpha_j P^{v}_j, 
		\quad 
		\alpha_j, P^{v}_j \in [0,1], 
		\quad 
		\sum \alpha_j = 1,
		\label{repequ}
	\end{equation}
	where each parameter $P^{v}_j$ (\eg, feedback score, task-specific quality indicator) is weighted by its importance coefficient $\alpha_j$.
	\paragraph*{Reputation Update} Global reputation scores evolve through a piecewise-weighted mean (PW-Mean) update rule:  
	\begin{equation}
		R^{v,i+1} =  
		\begin{cases}
			(1- \psi \mathsf{W_f}) R^{v,i} + \psi \mathsf{W_f} T^{v,i}, & T^{v,i} \geq T_{\theta}, \\[4pt]
			(1- \xi \mathsf{W_f}) R^{v,i} + \xi \mathsf{W_f} T^{v,i}, & T^{v,i} < T_{\theta},
		\end{cases}
		\label{equ:repmodel}
	\end{equation}

	where $R^{v,i}$ is the reputation of worker $v$ after its $i$-th interaction, $T^{v,i}$ is the corresponding trust value Eq.~\eqref{repequ}, $T_{\theta}$ is a trust threshold (\eg, $T_{\theta} = \overline{T}$), and $\mathsf{W_f}$ is the interaction weight Eq.~\eqref{equ1}. The coefficients $\psi$ and $\xi$ ($\xi > \psi$) control the asymmetry between positive and negative updates. This asymmetry penalizes malicious or low-quality interactions more than it rewards positive ones, making reputation harder to build than to lose.

	\quad  Eq.~\eqref{equ:repmodel} corresponds to a piecewise exponential moving average (EMA~\cite{das2011securedtrust}), which is a contraction mapping for $0 < \gamma \mathsf{W_f} < 1$, where $\gamma \in \{\psi,\xi\}$. This guarantees stability and exponential convergence toward the underlying trust value, as formally shown in Appendix~A. Larger values of $\gamma$ increase responsiveness but may amplify noise, whereas smaller values improve robustness at the cost of slower convergence. The asymmetry constraint $\xi > \psi$ enforces a security-oriented design in which negative interactions have a stronger impact than positive ones, limiting reputation inflation and improving resilience against collusion and opportunistic attacks. The trust threshold $T_{\theta}$ defines the minimum acceptable trust level. In practice, these parameters can be selected based on security requirements or dynamically adapted. For instance, the update coefficients can be dynamically adjusted as functions of the current reputation level or global system conditions, \eg,
	\begin{equation}
		\psi_i = \psi_0 \, R^{v,i}, 
		\quad
		\xi_i = \xi_0 \, (1 - R^{v,i}),
	\end{equation}
	which increases penalties for low-reputation workers while progressively stabilizing trusted ones. Similarly, the threshold $T_{\theta}$ can be defined using population statistics (\eg, mean or percentile of trust values) to reflect the current system state. This adaptivity preserves the convergence guarantees while improving robustness and generality across heterogeneous scenarios. Appendix~A provides concrete instantiations of Eq.~\eqref{repequ} and formally proves that the proposed update rule in Eq.\eqref{equ:repmodel} defines a bounded and stable dynamical system and guarantees exponential convergence toward the underlying trust level (Lemma 1–5 and Theorem 1).

	\subsubsection*{\textbf{Step 4. Mint a new Reputation Token}} 
	Now, if, after some interactions $k$, $u$ wants to use a different pseudonym to regain anonymity, let us call it $apk^{i+k}$, $u$ must mint a new token using $apk^{i}$ and the most recent reputation $R^{i+k}$. This is done following Protocol~\ref{protocol:mintrt}. If \texttt{TX\textsubscript{RM}} passes, the address public key $apk^{i+1}$ (not $apk^{i+k}$) is removed from the list $Valid$. The user cannot use it again and must spend or use the freshly minted token $\texttt{RT\textsuperscript{i+k}}$ for future interactions.

	\refstepcounter{protocol} 
	\begin{protocolbox}{$\mathcal{U} \xrightarrow[\text{}]{\texttt{MintRT}}$ $\mathbf{CSML}$ \{$RSC$\}}
		\label{protocol:mintrt}
		\begin{small}
			\begin{itemize}[leftmargin=0.5em]
				\item[\scriptsize{$\blacksquare$}] \textbf{Data:}   Address Key pair $(apk$, $ask)^{i}$
				\item[\scriptsize{$\blacksquare$}] \textbf{Result:} Mint a new \texttt{RT\textsuperscript{i+k}} with $R^{i+k} = R^{i}$.
			\end{itemize}
			\begin{enumerate}[leftmargin=0.5em]
				\item User \( u \) generates a new key pair \((apk, ask)^{i+k}\)
				\item Sample $s^{i+k}$ (serial number randomness)
				
				\item Set $cm_p^{i+k} := [\mathsf{COMM}_{r_1}(apk \parallel s)]^{i+k}$ with random $r_1^{i+k}$
				\item Set $cm_R^{i+k} := [\mathsf{COMM}_{r_2}(R \parallel cm_p)]^{i+k}$ with random $r_2^{i+k}$
				\item Form $\texttt{RT\textsuperscript{i+k}} := (apk, R, s, r_1, r_2, cm_R)^{i+k}$
				\item Form $\texttt{TX\textsubscript{RM}} := (R, cm_p, r_2, cm_R)^{i+k}$

				\item Invoke $RSC$: Submit \texttt{TX\textsubscript{RM}} to the CSML.\\
				- Accept only if $apk^{i} \in Valid$ and $cm_R^{i+k}$ is correct.
			\end{enumerate}
		\end{small}
	\end{protocolbox}

	\subsection{Anonymous Payment} \label{sec:$DARTC$pay}
	\quad Unlike traditional anonymous payment systems, which raise concerns about their use in money-laundering activities by cybercriminals and malicious actors, we use dynamic whitelisting via a smart contract to guarantee compliance. This means that only registered and authorized addresses can make payments on the CSML. To this end, we adapt the Tornado Cash protocol \cite{tornado} to our permissioned setting. Specifically, an access control smart contract ($ASC$) maintains the set of valid accounts. Only addresses $apk \in Valid$ are permitted to interact with the deposit smart contract ($DSC$). To deposit an amount $D$, a user (\eg, a requester) generates and signs a transaction \texttt{TX\textsubscript{deposit}} with the private key associated with their current reputation token \texttt{RT\textsuperscript{i}} (\ie, $apk^i$ $\in Valid$ ). To form \texttt{TX\textsubscript{deposit}}, the user first computes a commitment 
	\[
	cm_D = \mathsf{COMM}_r(s \parallel D),
	\] 
	by hashing a random secret $s$ concatenated with the amount $D$  using a trapdoor $r$. The signed transaction 
	\[
	\texttt{TX\textsubscript{deposit}} := (taskID, D, cm_D)
	\] 
	is then submitted to CSML, which inserts $cm_D$ into a Merkle tree of deposit commitments $DCTree$ and locks the amount $D$  in $DSC$. To prevent double-spending, each secret $s$ is deterministically mapped to a nullifier $n_s = \mathsf{NULL}(s)$ that is revealed at withdrawal.  
	
	The funds $D$  are transferred to the other party (\eg, workers) as payment after a successful interaction by invoking the withdraw function through a threshold signature scheme similar to the one detailed in Algo.~\ref{algReputation}.
	
	\quad In case of an unsuccessful interaction (\eg, timeout or invalid check), the requester can reclaim the funds locked in $DSC$ by withdrawing them to any valid address $apk$. To do so, the requester must produce a zk proof $\pi_D$ for the following: 
	\begin{quote}
		\textbf{``\textit{I know a valid opening $(s, r, D)$ s.t. $cm_D = \mathsf{COMM}_r(s \parallel D)$ appears as a leaf in a CRH-based Merkle Tree $DCTree$ whose root is $rt_D$}''}. 
	\end{quote}
	The proof $\pi_D$ is verified by $DSC$ as part of the \texttt{TX\textsubscript{withdraw}}. If $\pi_D$ is valid, the contract transfers the corresponding amount $D$  to the new account $apk$. The contract maintains a record of valid accounts and used nullifiers; therefore, if the target address $apk$ $\notin Valid$ or the nullifier $n_s$ has already been used, the withdrawal request is denied.

	\section{Theoretical Analysis} \label{sec:$DARTC$security} 
	
	\quad This section analyzes the security and privacy properties of $\mathsf{DARTIC}$ as defined in \S\ref{sec:$DARTC$threats}. We explicitly distinguish between: (i) \textit{component-level guarantees}, which are formally reduced to standard cryptographic assumptions (e.g., zkSNARK soundness, commitment binding, and collision resistance), and (ii) \textit{system-level security properties}, which emerge from the composition and interaction of these components under well-defined protocol interfaces. While network activities such as posting transactions may reveal identifiers such as IP addresses (mitigated by anonymous networks such as Tor~\cite{tor}), securing network-layer metadata is outside the scope of this analysis.
	
	\begin{itemize}
		\item \textit{Sybil resistance:} (system-level) In $\mathsf{DARTIC}$, each legitimate user is granted at most one valid credential per context. The IDML committee maintains a set $Granted_{ctx}$ of already-registered identifiers. If $M_{pk^u} \notin Granted_{ctx}$, a credential is issued; otherwise, it is rejected. This property is enforced through credential uniqueness enforced by the registration protocol and thus emerges from the composition of identity management and on-chain state tracking.

		\item \textit{Collusion resistance:} (system-level) Collusion resistance is achieved through the joint composition of (i) VRF-based randomness, (ii) threshold signing, and (iii) on-chain verification.
		
		\textit{- Randomness.} (component-level) This ensures that the selection of $k>t+1$ committee nodes ($cn_1, \dots, cn_{k}$) for verifying a user's claim is unpredictable. VRF~\cite{vrf} is used to generate a secure and verifiable random seed $r$. Any entity, such as IDML committee nodes, oracle nodes, or smart contracts, can verify the proof $\pi_{r}$ using the VRF public key. Using $r$ as a seed,  $k$ committee nodes are selected via a secure hash function. This tamper-proof process ensures the prover and verifier cannot manipulate node selection, and collusion requires compromising a significant and randomly chosen fraction, which is computationally infeasible.
		
		\textit{- Distributed Signing and Threshold Security.} (component-level)
		Distributed signing requires independent validation from multiple committee nodes. During credential issuance (\S\ref{subSec:access}), each node verifies the proof $\Pi$ and, if valid, generates a partial signature $\sigma_{i} = \text{Sig}_{sk_i}(\Phi)$. The DON aggregates these partial signatures into a final signature $\sigma_{cred}$  only if at least $t+1$ valid signatures are collected. Therefore, even if some nodes collude with the prover, they cannot forge the final signature unless $t+1$ nodes are compromised.
		
		\textit{- On-Chain Verification.} (system-level) The blockchain enforces correct aggregation and prevents unauthorized credential issuance. It guarantees transparency by publicly verifying aggregated signatures through the $ISC$ and $RSC$.

		\item \textit{User Privacy:} (component-level $\rightarrow$ system-level) Privacy follows from: (i) hiding commitments, (ii) zero-knowledge proofs in zkSNARKs, and (iii) MPC-based privacy guarantees in DECO~\cite{deco}. Since no plaintext identity information is used during credential issuance or interaction evaluation, the composition of these primitives ensures that user identity is not revealed at the system level.

		\item \textit{Unlinkability:} (System-level) Unlinkability is obtained by composing zkSNARK-based proofs with commitment schemes and Merkle-set membership proofs (\S\ref{sec:bblocks}).
		
		\textit{- User--Pseudo unlinkability.}  (component-level guarantee)
		Each user \( u \) generates a unique key pair \((sk_u, pk_u)\) when minting a token AT or RT. The public key $pk_u$ is a pseudonym in the system. The zkSNARK \(\mathsf{Prov}\) algorithm allow \( u \) to spend a token (AT or RT) without revealing the corresponding private key \( sk_u \) or linking the current pseudonym to any previous pseudonym.  Formally, following Definitions~\ref{def1} and~\ref{def2}: \begin{inparaenum}[(i)]
			\item to spend an AT with \( cm_A = \mathsf{COMM}_r(sk_u) \), user \( u \) provides a zkSNARK proof that \( cm_A \) exists as a leaf in \( ACTree \) without revealing \( sk_u \), $r$ or any other leaves.  
			\item Similarly, to spend an RT, \( cm_R^i \) is proven to exist in \( RCTree \).  
		\end{inparaenum}
		zkSNARK zero-knowledge ensures that spending a token does not reveal the underlying secret key or link it to prior pseudonyms.

		\textit{- Pseudonym--pseudonym unlinkability (system-level).} Fresh key generation combined with zkSet membership proofs prevents correlation across transactions. The adversary's advantage remains negligible under collision resistance and zkSNARK soundness. 
		Unlinkability between different pseudonyms of the same user is achieved as follows: \begin{inparaenum}[(i)]
			\item Each AT or RT issuance generates a new key pair and associated commitment. These commitments are inserted into their respective Merkle trees, \( ACTree \) or \( RCTree \), 
			\item When spending a token, the zkSet proof proves membership in the tree but does not reveal linkage. Because the Merkle proofs are combined with zkSNARKs (\S\ref{sec:bblocks}), an adversary cannot distinguish whether two different pseudonyms belong to the same user. The adversary’s advantage is bounded by \( \epsilon \), which is negligible under standard cryptographic assumptions (collision-resistance of hash functions and soundness of zkSNARKs).  
		\end{inparaenum}
		
		\item \textit{Accountability:} (system-level) Accountability arises from the interaction of reputation updates, AT-binding constraints, and enforcement via credential revocation: (i) misbehavior sharply reduces a user’s reputation, limiting influence and access; (ii) users below a reputation threshold are sanctioned (\ie, removed from $Valid$), preventing interactions under their current pseudonym; (iii) re-entry requires minting a new AT, constrained by the \textit{one-AT-per-context} rule (AT binding), preventing circumvention of sanctions.
		
		\item \textit{Reputation Binding:} (component-level $\rightarrow$ system-level property) This relies on ECDSA unforgeability (under Elliptic Curve Discrete Logarithm Problem (ECDLP) hardness~\cite{gennaro2020one, gennaro2018fast}), commitment binding, and zkSNARK soundness. These primitives ensure that a committed value cannot be altered and that proofs cannot be generated for inconsistent values, thereby maintaining a cryptographically fixed link between identity and reputation.
		
		\item \textit{Forward Reputation Binding:} (system-levely) This property holds under the composition of (i) signature unforgeability, (ii) commitment binding/hiding, and (iii) zkSNARK soundness. Together, they ensure that reputation evolution remains consistent, private (if not shown), and correctly bound to the corresponding entity. A formal analysis of zkSet circuit correctness is provided in Appendix~B.

	\end{itemize}
	
	\color{black}
	
	\section{Evaluation and Results} \label{sec:$DARTC$evaluation}
	\quad  In this section, we present an experimental evaluation of $\mathsf{DARTIC}$ to assess its efficiency, scalability, and resilience. Additional experiments that compare the proposed PW-Mean model with baseline models and show the optimal RT reuse window are provided in Appendix~C.
	\subsection{Experimental Setup}
	\quad We developed a proof of concept for $\mathsf{DARTIC}$ and the proposed protocols. The circuits are implemented using the $\mathsf{circom}$ domain-specific language (DSL) and the $\mathsf{circomlib}$ library\footnote{\href{https://github.com/iden3/circomlib}{https://github.com/iden3/circomlib}}, while $\mathsf{snarkjs}$\footnote{\href{https://github.com/iden3/snarkjs}{https://github.com/iden3/snarkjs}} is used to compile the circuits and perform the powers of tau ceremony for the trusted setup \cite{ptau}. The smart contracts for $\mathsf{DARTIC}$ are coded in $\mathsf{Solidity}$. For $ACTree$, $RCTree$, and $DCTree$, we implement a 20-level Merkle tree using MiMCSponge\footnote{\href{https://github.com/iden3/circomlib/blob/master/circuits/mimcsponge.circom}{https://github.com/iden3/circomlib/blob/master/circuits/mimcsponge.circom}}, a lightweight hash function optimized for ZKPs. Each tree operation requires 40 MiMCSponge calls—one hash per level, with two calls per hash to compute the node (\textit{\_left}) and its sibling (\textit{\_right}).
	
	\quad \textit{Datasets and models.} To evaluate the effectiveness and robustness of the proposed reputation model and the PW-Mean aggregation method, we conducted experiments that simulate federated learning and crowdsensing scenarios. We select datasets to serve as realistic sources of heterogeneous, noisy, and potentially adversarial interaction outcomes, rather than as benchmarks for learning performance. Specifically:  
	\begin{inparaenum}[(i)]
		\item For federated learning tasks, we use MNIST~\cite{LeCun2005TheMD} and CIFAR-10~\cite{krizhevsky2009learning} to train a simple CNN and a ResNet-18 model, respectively. These datasets generate controlled yet diverse model updates, allowing us to emulate honest, low-quality, and malicious contributions.  
		\item For crowdsensing tasks, we use the OpenSense Zurich dataset~\cite{li2012sensing}, which provides realistic environmental sensing traces. The dataset allows us to simulate varying quality of service, data noise, and strategic misreporting by participants.  
		\item Interaction outcomes from all datasets are mapped to quality signals, which serve as inputs to the PW-Mean reputation update function. 
	\end{inparaenum}  
	Thus, these datasets are used as behavioral generators to model service reliability and adversarial dynamics, consistent with previous work on trust and quality evaluation~\cite{koutsos2024mathsf, zebralancer, duan2019aggregating}.

	\quad \textit{Environment.} Experiments are conducted on a cluster of two HPE ProLiant XL225n Gen10 Plus servers. Each server is equipped with dual AMD EPYC 7713 64-core 2.0 GHz processors and 2 × 256 GB of RAM. Federated learning tasks are executed on a GPU card ``NVIDIA A6000''.
	
	
	\vspace{-0.4cm}
	\subsection{Execution Time}
	\quad We evaluate the zkSNARK constructions in $\mathsf{DARTIC}$ using two proving systems: Groth16~\cite{groth} and PlonK~\cite{plonk}. Groth16 is a circuit-specific SNARK that requires a structured reference string (SRS) per circuit but offers short constant-size proofs and fast verification. Conversely, PlonK is a universal SNARK scheme with a single updatable SRS for all circuits, albeit with higher proof generation costs and longer call data.

	\begin{table}[t]
		\centering
		\caption{Execution time (s) of individual zkSNARK proofs Groth16~\cite{groth} vs. PlonK~\cite{plonk} proving systems.}
		\setlength{\tabcolsep}{3pt} 
		\renewcommand{\arraystretch}{1.2} 
		\resizebox{0.46\textwidth}{!}{ 
			\begin{tabular}{cccccc}
				\hline
				\textbf{Proof} & \bf System & \bf Setup  & \bf  $\textbf{Prove}$ & \bf $\textbf{Verify*}$ &  \bf Call Data size   \\   \hline 
				
				\multirow{2}{*}{$\pi_{AS}$}  &  
				$\mathsf{Groth16}$ & 150$\pm$5 &\:\: 2.40$\pm$0.10 & \:\: 0.73$\pm$0.05 & \:\:  705B  \\ 
				& $\mathsf{PlonK}$ & 90$\pm$5 & \:\: 67.00$\pm$0.5 & \:\: 0.76$\pm$0.05 & \:\:  1750B     \\ \hline
				
				\multirow{2}{*}{$\pi_{RS}$}  &  
				$\mathsf{Groth16}$ & 150$\pm$5 & \:\: 2.90$\pm$0.10 & \:\: 0.95$\pm$0.10 & \:\:  705B  \\ 
				& $\mathsf{PlonK}$ & 90$\pm$5 & \:\: 79.5$\pm$0.50 & \:\: 0.93$\pm$0.10 & \:\:  1750B      \\ \hline
				
				\multirow{2}{*}{$\pi_I$}  &  
				$\mathsf{Groth16}$ & 30$\pm$5 & \:\: 0.48$\pm$0.05 & \:\: 0.64$\pm$0.05 & \:\:  705B  \\
				& $\mathsf{PlonK}$ & 15$\pm$5 & \:\: 3.40$\pm$0.10 & \:\: 0.67$\pm$0.05 &  \:\:  1750B  \\ \hline
				
				\multirow{2}{*}{$\pi_{D}$}  &  
				$\mathsf{Groth16}$ & 150$\pm$5 & \:\: 2.30$\pm$0.10 & \:\: 0.71$\pm$0.05 & \:\:  705B  \\ 
				& $\mathsf{PlonK}$ & 90$\pm$5 & \:\: 65.0$\pm$0.50 & \:\: 0.73$\pm$0.05 & \:\:  1750B      \\ \hline
				
			\end{tabular}
		}
		\label{$DARTC$:tab1}
		\begin{tablenotes}
			\footnotesize
			\item $\textbf{*}$on-chain verification time 
		\end{tablenotes}
	\end{table}
	
	\quad Table~\ref{$DARTC$:tab1} presents benchmark results for proof generation and verification times across four circuits: $\pi_{AS}$ (access token), $\pi_{RS}$ (reputation token), $\pi_{I}$ (interaction proof), and $\pi_{D}$ (deposit proof). For each proof, we report timings under both Groth16 and PlonK. On average, Groth16 enables efficient proving within $0.50$--$2.90$s and verifying within $0.60$--$0.95$s per proof, while PlonK requires $3.40$s to over $80$s for proof generation, depending on circuit complexity.
	
	\quad Regarding the trusted setup, Groth16 requires circuit-specific SRS generation involving heavy pairing operations. While this introduces setup overhead, $\mathsf{DARTIC}$ uses a fixed and limited number of circuits (four), making Groth16's per-circuit setup a one-time cost that does not impede scalability. In contrast, PlonK decouples setup from circuit structure, relying on polynomial commitment schemes that facilitate faster universal SRS generation. However, due to its significantly higher proving time and proof size (1.75KB vs. 705B), we find Groth16 more favorable for our current architecture.

	\begin{figure*}[t]
		\centering
		\hspace{-0.2in}
		\subfloat[mintAT (\texttt{TX\textsubscript{AM}})]{
			\includegraphics[scale=0.22]{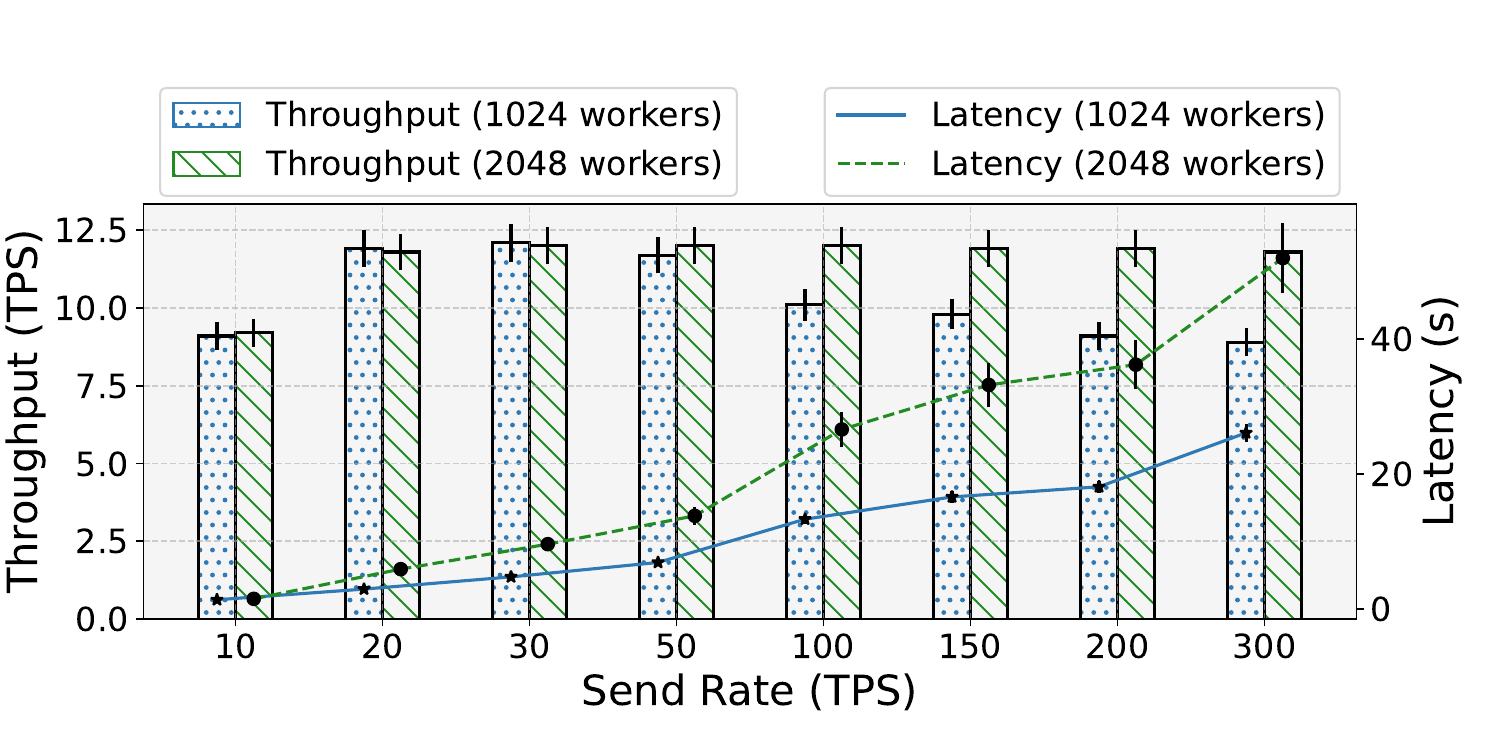}
			\label{fig3a}
		}
		\subfloat[mintRT (\texttt{TX\textsubscript{RM}})]{
			\hspace{-0.2in}
			\includegraphics[scale=0.22]{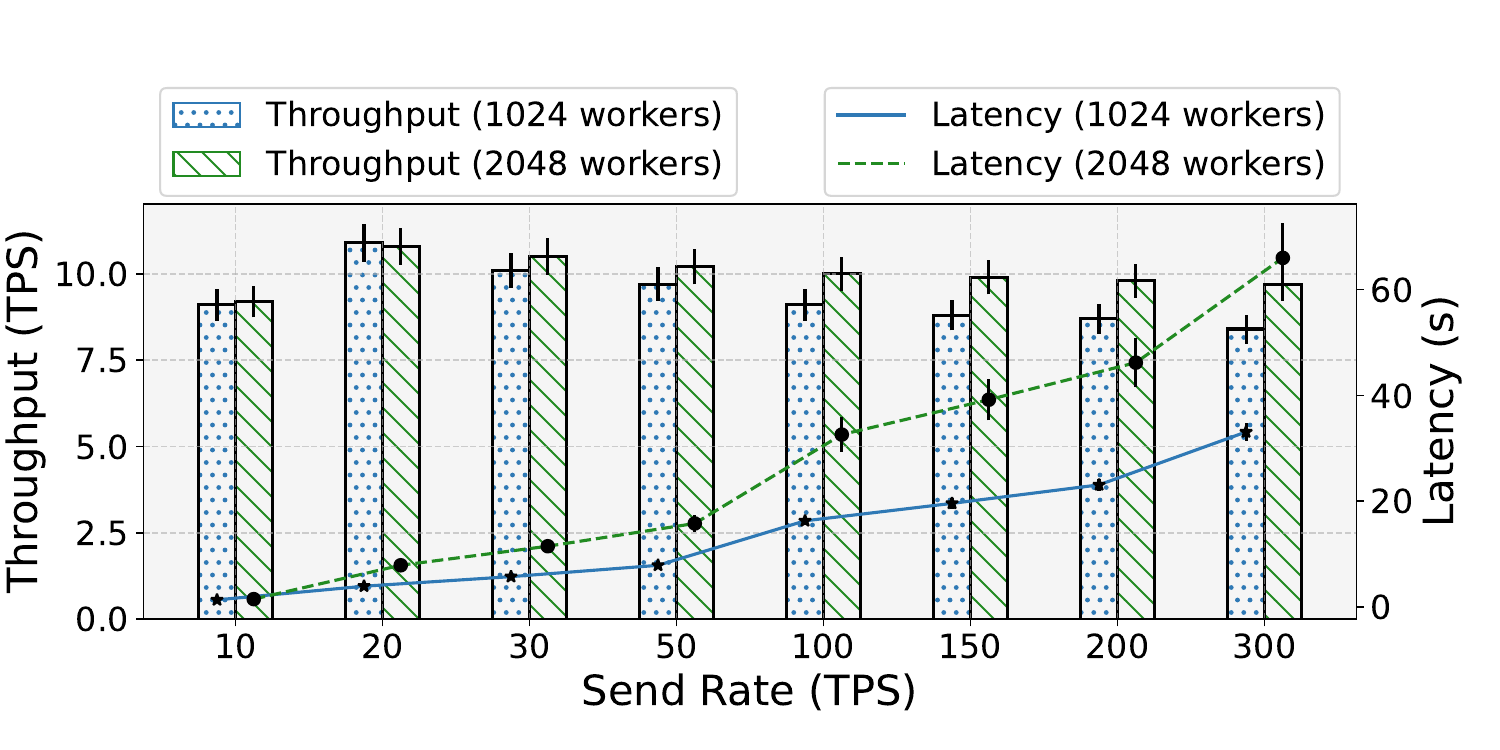}
			\label{fig3b}
		}
		\subfloat[useRT (\texttt{TX\textsubscript{RU}})]{
			\hspace{-0.2in}
			\includegraphics[scale=0.22]{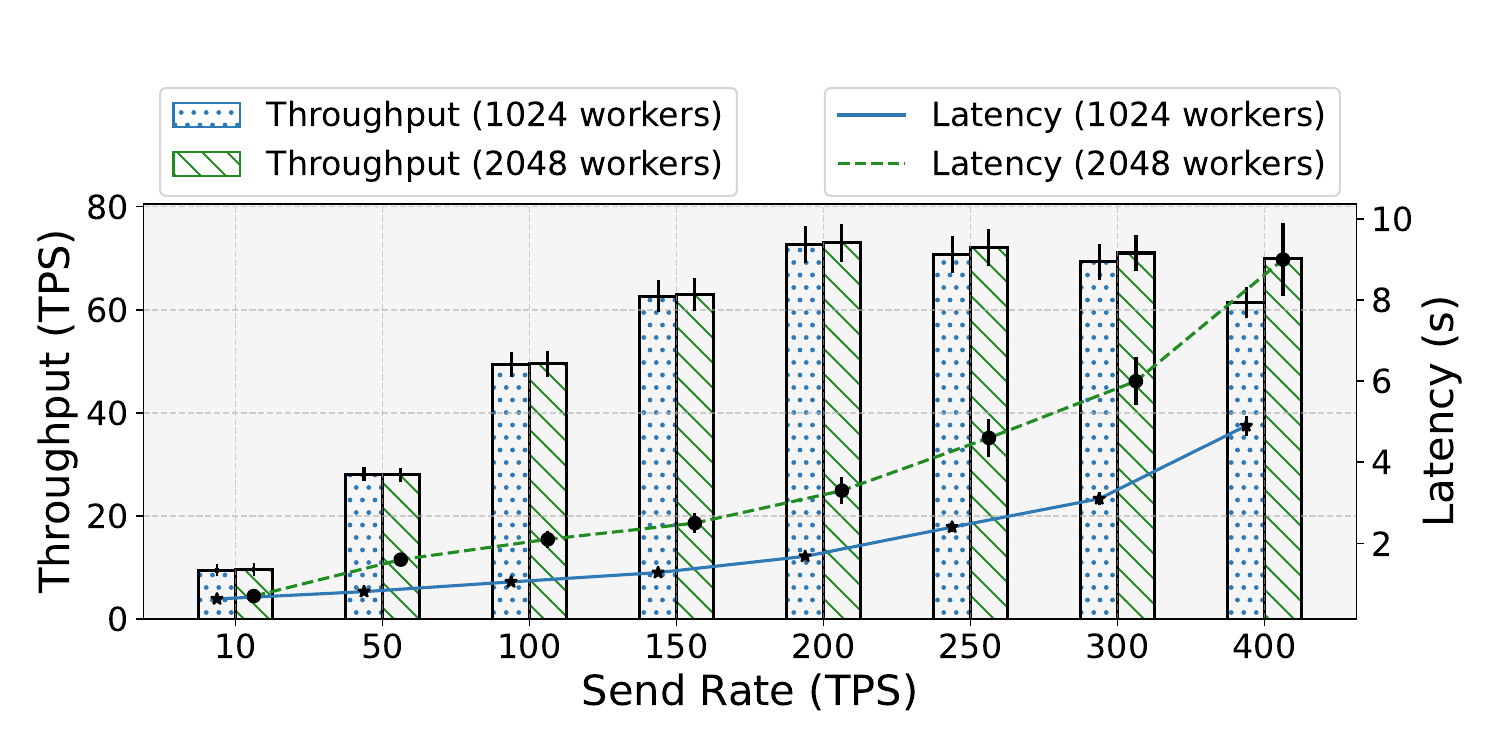}
			\label{fig3c}
		}
		
		\subfloat[updateRT (\texttt{TX\textsubscript{upRep}})]{
			\hspace{-0.22in}
			\includegraphics[scale=0.22]{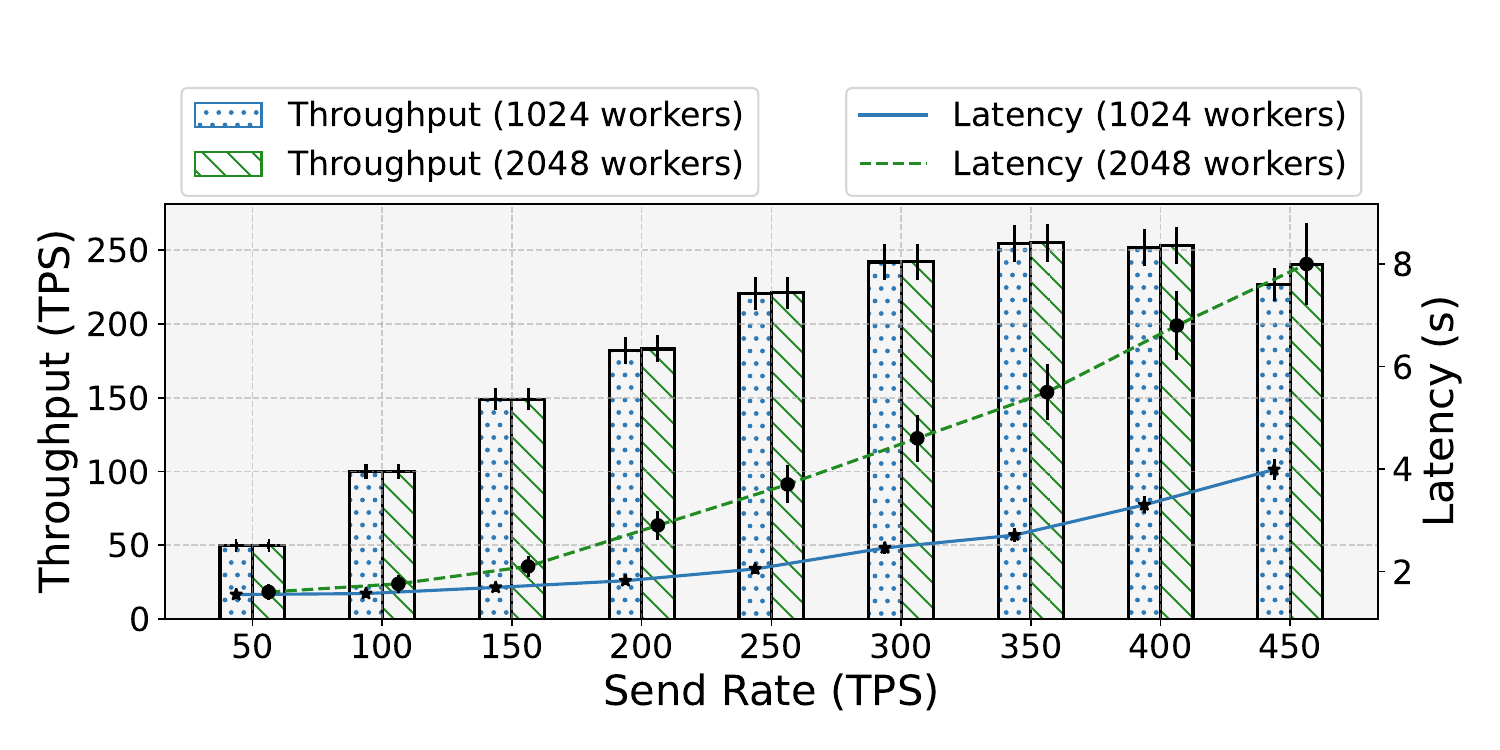}
			\label{fig3d}
		}
		\subfloat[deposit (\texttt{TX\textsubscript{deposit}})]{
			\hspace{-0.22in}
			\includegraphics[scale=0.22]{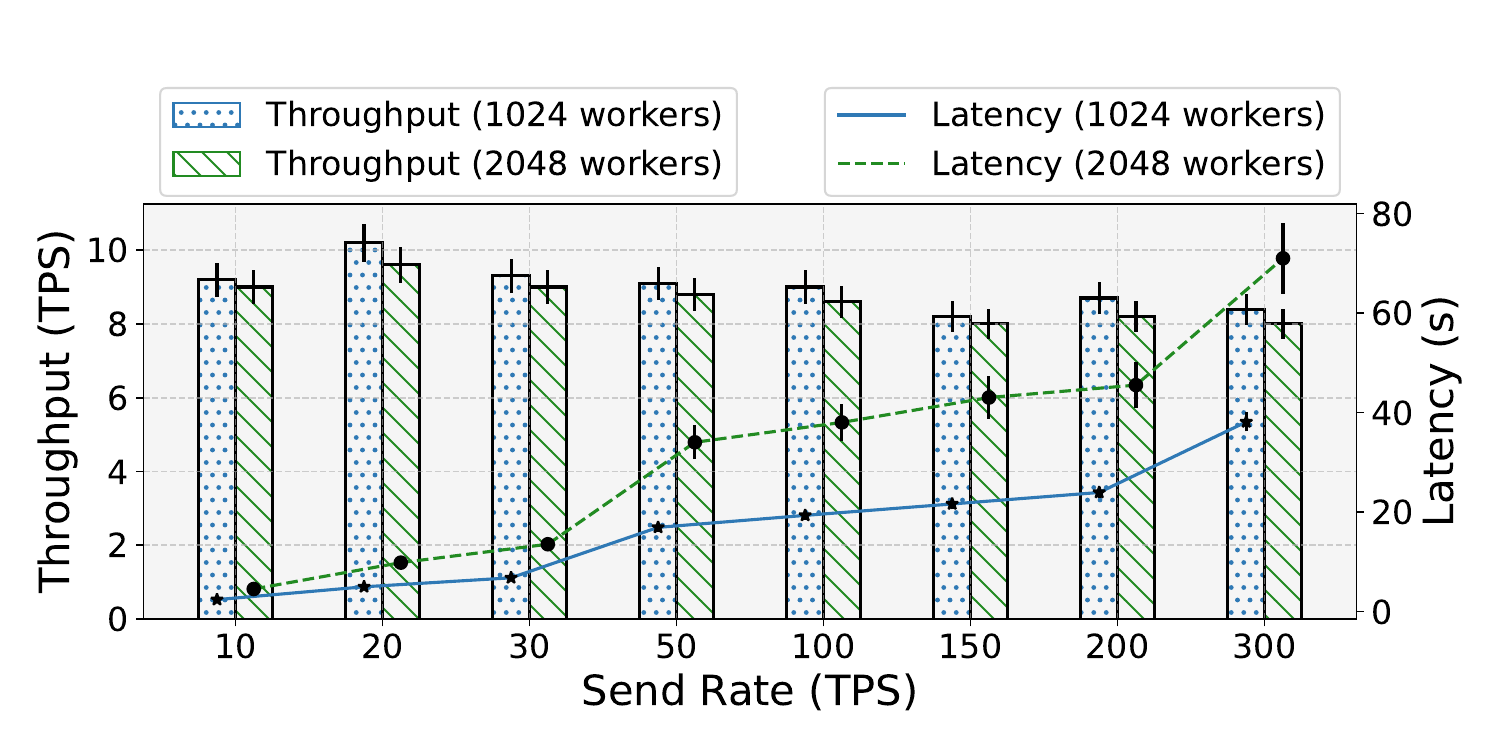}
			\label{fig3e}
		}
		\subfloat[withdraw (\texttt{TX\textsubscript{withdraw}})]{
			\hspace{-0.22in}
			\includegraphics[scale=0.22]{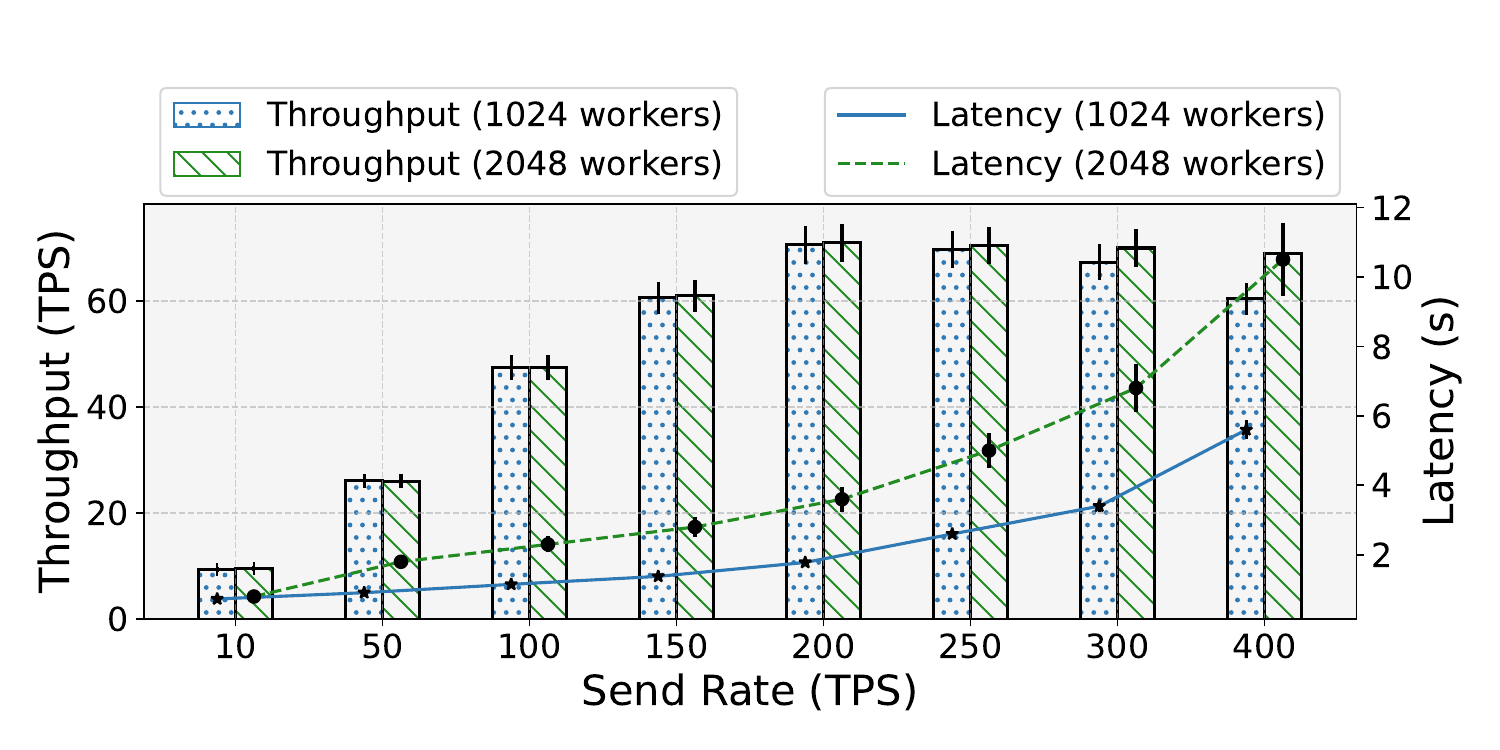}
			\label{fig3f}
		}
		\caption{Latency and Throughput of $\mathsf{DARTIC}$.}
		\label{fig3}
	\end{figure*}

	\begin{figure}[t]
		\centering
		\includegraphics[width=0.9\linewidth, height=.45\linewidth]{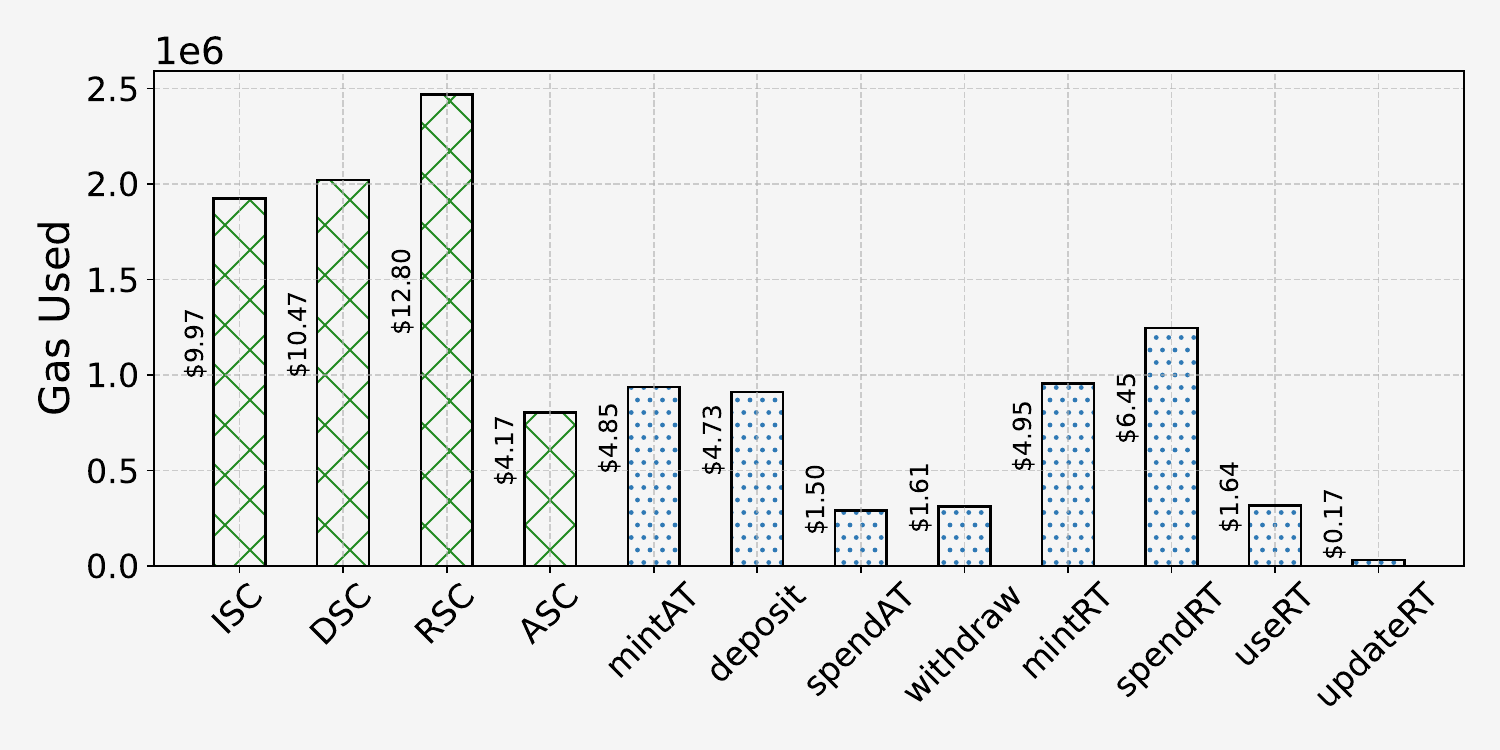}
		\caption{Deployment and invocation gas costs. $
			\text{Cost}_{\text{USD}} = \text{gas}_{\text{used}} \times \text{gas}_{\text{price}}^{\text{gwei}} \times 10^{-9} \times \text{ETH}_{\text{price}}^{\text{USD}}
			$.
		}
		\label{fig:gascost}
	\end{figure} 
	\vspace{-0.4cm}
	\subsection{L1 Performance} 
	\quad We evaluate $\mathsf{DARTIC}$ on-chain performance on a single-layered (L1) blockchain using Caliper\footnote{\href{https://github.com/hyperledger/caliper-benchmarks}{https://github.com/hyperledger/caliper-benchmarks}}. To this end, we deploy and benchmark our smart contracts on a local Hyperledger Besu\footnote{\href{https://besu.hyperledger.org}{https://besu.hyperledger.org}} network comprising 15 validators using Quorum-BFT as the consensus protocol. The experiments involve using two workloads (1024 and 2048 workers) while altering the TX sending rate (ranging from 10 to 450 TPS) using a consistent network configuration (standard block time =1s and 12 validators) for the main operations. The average values of latency and throughput are shown in Fig.~\ref{fig3} and the respective gas costs are presented in Fig.~\ref{fig:gascost}.
	
	\quad The system throughput scales proportionally with the transaction sending rate until it approaches its saturation point. For instance, \texttt{TX\textsubscript{upRep}} reaches a peak of $255$ TPS at a send rate of $350$ TPS, after which performance begins to degrade. The system maintains a low end-to-end latency (below $3$s) for \texttt{TX\textsubscript{upRep}} as long as the workload remains within this optimal range, indicating efficient handling before overload conditions occur. The other transaction types exhibit a similar pattern but achieve lower peak throughput due to their higher computational complexity. In particular, \texttt{TX\textsubscript{AM}}, \texttt{TX\textsubscript{RM}}, and \texttt{TX\textsubscript{deposit}} are more resource-intensive because they require inserting the $cm_A$, $cm_R$, and $cm_D$ commitments into the respective $ACTree$, $RCTree$, and $DCTree$ Merkle structures. This involves recalculating the Merkle root by recomputing all hash values along the path from the newly added leaf to the root, which introduces additional overhead. 
	
	\quad L1 results confirm the framework is practical for small to medium (up to 2048 active workers) platforms, though $\mathsf{DARTIC}$ performance can be boosted via scaling mechanisms. The next section shows how these address L1’s inherent limits.

	\begin{table}[t]
		\centering
		\caption{Comparison of scaling techniques: Baseline (Groth16) vs. Proof aggregation (SnarkPack)  vs. L2 batching.}
		\setlength{\tabcolsep}{3pt}
		\renewcommand{\arraystretch}{1.2}
		\resizebox{\columnwidth}{!}{
			\begin{tabular}{lccc}
				\hline
				\textbf{Metric} & \textbf{Groth16} & \textbf{SnarkPack} & \textbf{zkRollups} \\ \hline
				
				Verifier Calls         & $n$            & 1                & 1 (batch $n$) \\
				Proof Size (Call Data) & $n \cdot 705$B & $\sim$800--1000B & $\sim$1--2KB total \\
				Verification Time      & $n \cdot 700$--950ms & $\sim$1.0--1.3s & $\sim$1.5--2.0s \\
				Proving Time           & $t_{prov}$ per proof & $t_{prov}+\delta_1$ & $t_{prov}+\delta_2$ \\
				Gas Cost               & $n \cdot G$    & $\sim$1.3G       & $\ll G$ per tx \\
				Scalability            & Low ($n$ on-chain) & Moderate (100s) & High (1000s) \\
				Privacy Guarantee      & Maintained & Maintained & Maintained \\
				\hline
			\end{tabular}
		}
		\begin{tablenotes}
			\footnotesize
			\item[] $t_{prov}$ Parallel client-side proving.
			\item[] \(\delta_2\) the additional operator-side batching overhead.
			\item[c] “Maintained” client-side proving.
		\end{tablenotes}
		\label{tab:batching}
	\end{table}
	\vspace{-0.4cm}
	\subsection{Scalability} \label{sub:scalability}
	
	\textbf{1. Beyond individual proofs.} While Groth16 is efficient for individual statements, large-scale deployment of $\mathsf{DARTIC}$—with potentially thousands of token validations per epoch—demands scalable verification. To this end, we investigate proof \emph{aggregation} and L2 \emph{batching} via two techniques: \textit{SnarkPack}~\cite{snarkpack} and \textit{zkRollups}~\cite{zkrollups}. Table~\ref{tab:batching} compares these.
	
	\quad \textbf{SnarkPack}~\cite{snarkpack} aggregates multiple proofs into a single succinct proof. This is done off-chain with Groth16 compatibility. The aggregated verifier runs only once, with slightly increased verification time ($\sim$1.3s for up to 100s proofs) and reduced calldata size. This results in a reduced gas cost (87\%, n = 10), which is important for use cases where a user (\ie, an aggregator) submits multiple proofs in a single session.
	
	\quad \textbf{zkRollups}~\cite{zkrollups} batch multiple transactions into a single proof using PlonK~\cite{plonk}. Here, proving occurs in two layers:  
	\begin{itemize}  
		\item \textit{Client-side proving}: Each user independently generates a Groth16 proof of their transaction locally, incurring a proving time $t_{prov}$ per transaction. This step is fully parallelized and distributed across users.  
		\item \textit{Operator batch proving}: The zkRollup operator aggregates the transaction state transitions into a single succinct proof using an efficient PlonK, incurring an additional overhead $\delta_2$. This overhead is amortized over the entire batch.  
	\end{itemize}
	\quad Although the aggregation time remains comparable to or slightly higher than the baseline, parallelizing user proofs and amortizing the batch proving overhead with zkRollups drastically reduces the \emph{wall-clock}. On the verification side, zkRollups require only a single on-chain verifier call, regardless of batch size. While the total verification time ($\sim$1.5–2.0s) is slightly higher than SnarkPack, it remains constant (1 proof per batch) with respect to $n$. This efficiency also translates into significantly lower gas costs—well below $G$.

	\begin{figure}
		\centering
		\includegraphics[width=0.95\linewidth]{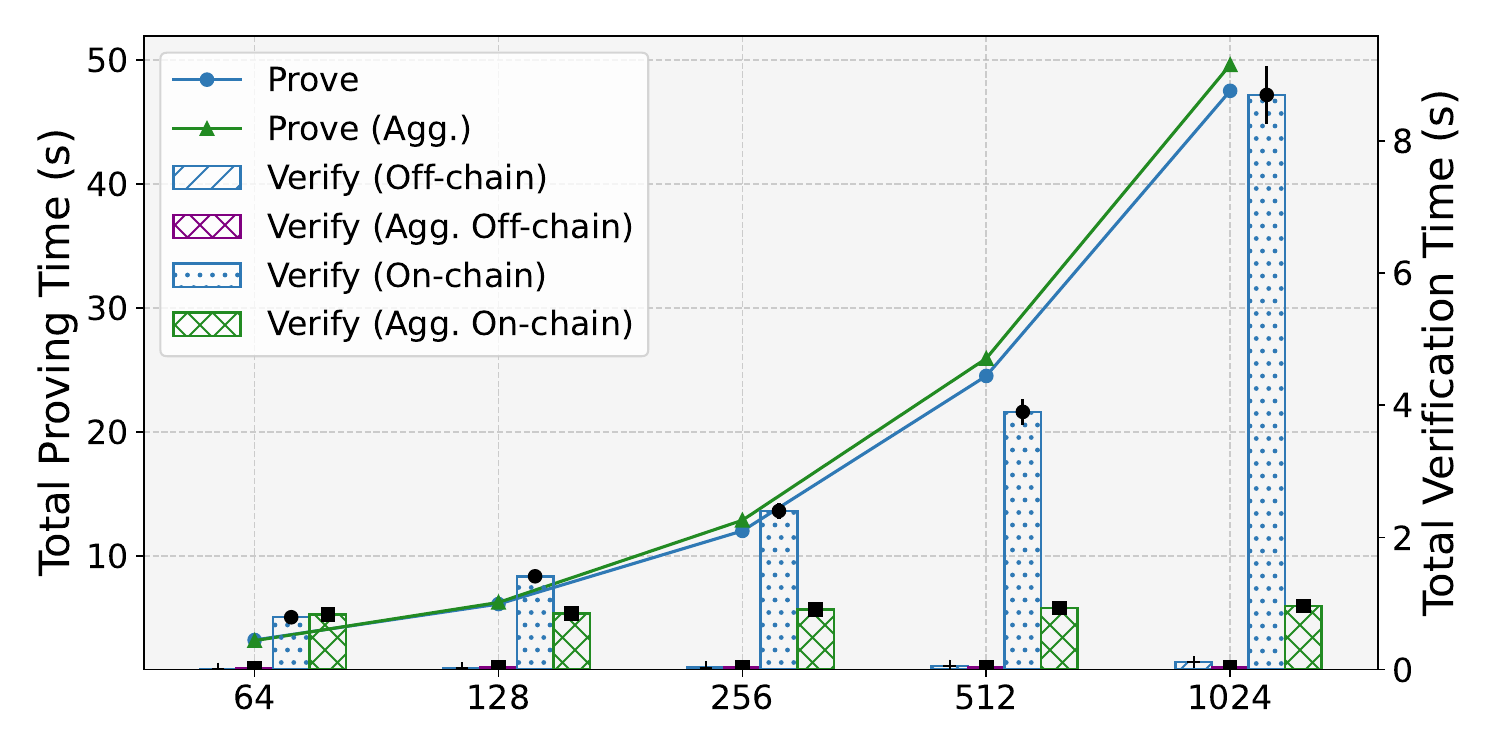}
		\caption{Proving and verification times of \texttt{spendRT} vs. number of workers.}
		\label{fig:prove_veriy_compare}
	\end{figure}

	\textbf{2. Impact of Aggregation.} We evaluate $\mathsf{DARTIC}$ with and without proof aggregation over 64–1024 workers, reporting \emph{proving} and \emph{verification} times of $\pi_{RS}$ (\texttt{SpendRT}), shown in Fig.~\ref{fig:prove_veriy_compare}. On the proving side, parallel execution across 64 workers exhibits near-linear scalability, with runtimes ranging from $3.2$s (64-batch) to $47.5$s (1024-batch). With aggregation, the same execution introduces only marginal overhead, primarily due to the individual proof checks ($<0.112$s) and the generation of the aggregation proof. For the largest batch, this results in $49.6$s, which remains within $5\%$ of the baseline. Verification, on the other side, exhibits a sharper pattern. In the absence of aggregation, the system can accommodate at most $80$ proof verifications per second (1-second block), yielding verification latencies that grow from $0.84$s (64 proofs) to $8.7$s (1024 proofs). By contrast, on-chain verification of aggregated proofs achieves logarithmic scaling, with a latency bounded between $0.84$s and $0.96$s across all batch sizes. As a result, aggregation is nearly cost-free for proving and cuts on-chain verification by an order of magnitude.

	\begin{figure}
		\centering
		\includegraphics[width=1.0\linewidth]{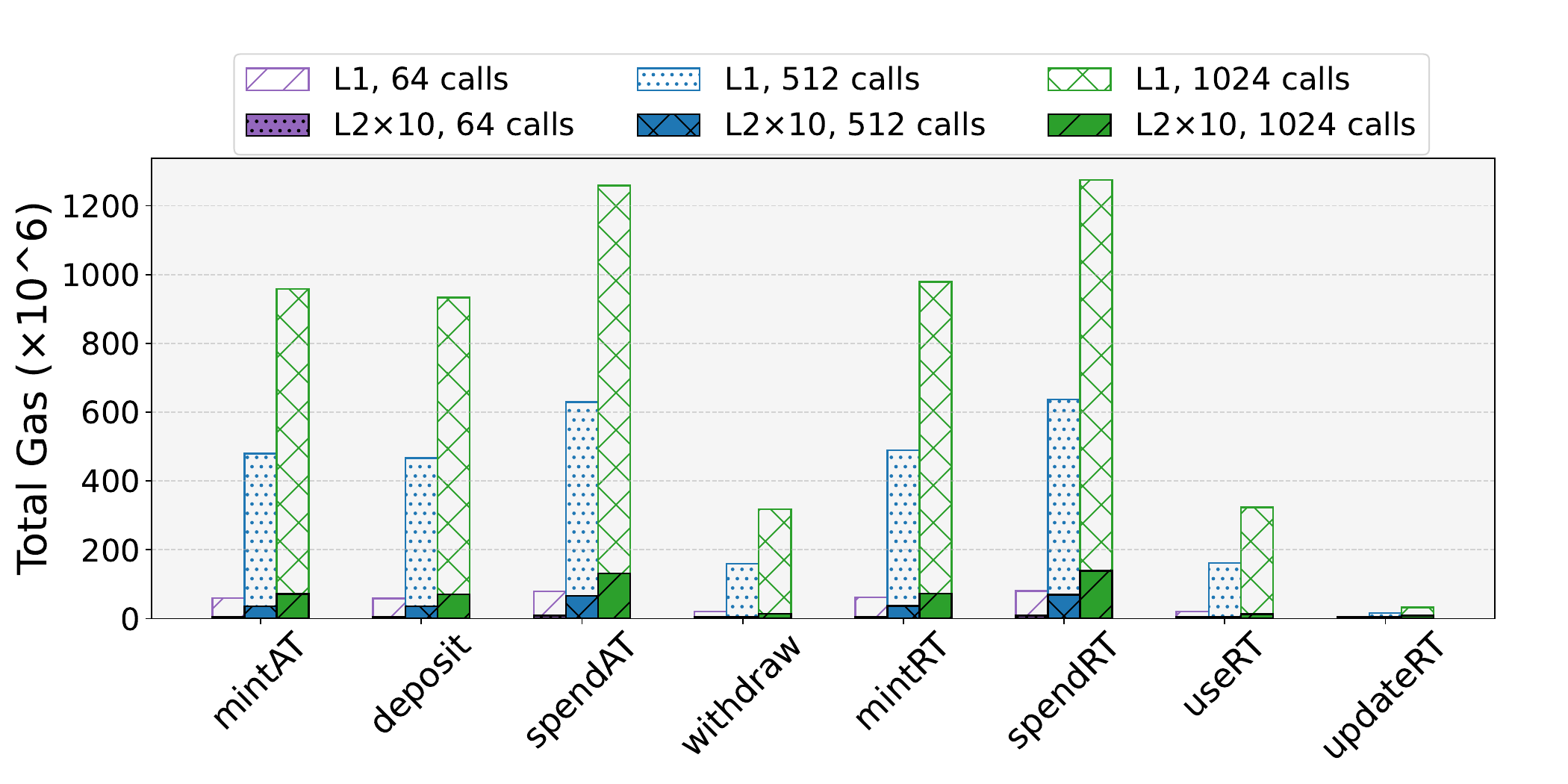}
		\caption{Total Gas Cost: L1 vs L2 (L2×10 for visibility).}
		\label{fig:l1_l2_gas}
	\end{figure}

	\textbf{3. Gas Efficiency of L2 vs L1.}
	We now evaluate the gas consumption of key system functions on Layer 2 (zkSync\footnote{\href{https://docs.zksync.io/zksync-era}{https://docs.zksync.io/zksync-era}}) relative to Layer 1. Each L1 call aggregates multiple L2 batches, with the maximum aggregation factor $m$ determined by function complexity and L1 gas limits.
	
	\begin{itemize}
		\item Heavy functions (\texttt{spendAT}, \texttt{spendRT}): $m=2$
		\item Medium functions (\texttt{mintAT}, \texttt{mintRT}, \texttt{deposit}): $m=3$
		\item Light functions (\texttt{withdraw}, \texttt{useRT}, \texttt{updateRT}): $m=8$
	\end{itemize}
	The \textit{total gas cost on L2} is computed as:
	\begin{multline}    
		\label{eq:L2_total}
		G_{\text{total,L2}} = \lceil \frac{\#\text{batches}}{m} \rceil \cdot G_{\text{commit}} + \lceil \frac{\#\text{batches}}{m} \rceil \cdot G_{\text{verify}} \\
		+ \lceil \frac{\#\text{batches}}{m} \rceil \cdot G_{\text{execute}}
	\end{multline}
	where $G_{\text{commit}}, G_{\text{verify}}, G_{\text{execute}}$ denote the base gas cost of a single aggregated call handling up to $m$ batches. The ceiling function $\lceil \cdot \rceil$ accounts for partially filled aggregated calls.
	L1 total gas cost scales linearly with the total number of calls:
	\begin{equation}
		\label{eq:L1_total}
		G_{\text{total,L1}} = \#\text{total\_calls} \cdot G_{\text{L1}}
	\end{equation}

	\quad Fig.~\ref{fig:l1_l2_gas} summarizes the results with detailed gas values provided in Appendix~C Table~I. L2 execution drastically reduces gas usage across all functions. Functions with cryptographic overhead (\eg, \texttt{spendRT}) achieve $96\times$ improvement. Medium functions achieve over $130\times$ reduction due to aggregation. Lightweight functions show improvements from $5\times$ (small batch counts) to over $230\times$ (large workloads). This shows that $\mathsf{DARTIC}$ scales efficiently with L2 zk-batching, supporting hundreds of operations per L1 call while preserving verifiability and security.
	
	\vspace{-0.4cm}
	\subsection{On-Chain Cost and end-to-end (E2E)  Latency}
	We compare the on-chain costs and latency with two representative privacy frameworks with comparable decentralized settings: AVeCQ~\cite{koutsos2024mathsf}, Duaan et al.~\cite{duan2019aggregating}, focusing on their E2E efficiency and scalability with respect to the number of workers. Table~\ref{tab:cost_comparison} summarizes the findings. The on-chain cost for $\mathsf{DARTIC}$ depends on the number of workers \(N\) and is composed of several contributions: Task creation: \texttt{newTask} = $289{,}070$ gas; Participation: \texttt{subToTask} = \(179{,}180 \times N\) gas; Worker selection: $33{,}200$ gas; Reputation update: \texttt{updateRT} = \(31{,}950 \times N\) gas; and Reputation token spending (optimal window = 5): \(\frac{1}{5} \cdot 1{,}244{,}750 \times N = 248,950 \times N\) gas. This yields a total gas cost without aggregation: $\text{Gas}_{\text{$\mathsf{DARTIC}$}} = 322{,}270 + 460{,}080 \cdot N$ gas. With SnarkPack aggregation (64 proofs per on-chain verification for both \texttt{subToTask} and \texttt{spendRT}), the gas cost is significantly reduced: $\text{Gas}_{\text{$\mathsf{DARTIC}$}}^{\text{agg}} = 322{,}270 + 38{,}640 \cdot N \quad \text{gas}$.
	
	\begin{table}[t]
		\centering
		\caption{On-Chain Cost and E2E  Latency Comparison. $\text{Cost (Wei)} = \text{Gas} \times 5 \times 10^9$.}
		\label{tab:cost_comparison}
		\begin{threeparttable}
			\begin{tabular}{lccc}
				\hline
				\textbf{System} & \textbf{Workers} & \textbf{E2E Latency} & \textbf{On-Chain Cost} \\
				\hline
				AVeCQ~\cite{koutsos2024mathsf} (Image)& 39 & $<$8mins* & $<$19 MWei \\
				AVeCQ~\cite{koutsos2024mathsf} (Review) & 128 & $<$18mins* & 55 MWei \\
				Duan et al.~\cite{duan2019aggregating} & 128 & 30mins* & $>$416 MWei \\
				$\mathsf{DARTIC}$ (no agg.) & 39 & $<$3mins & $<$91 MWei \\
				$\mathsf{DARTIC}$ (no agg.) & 128 & $<$5mins & $<$263 MWei \\
				$\mathsf{DARTIC}$ (64 proofs) & 39 & $<$2mins & $<$9 MWei \\
				$\mathsf{DARTIC}$ (64 proofs) & 128 & $<$2mins & $<$26 MWei \\
				\hline
			\end{tabular}
			
			\begin{tablenotes}
				\footnotesize
				\item * Denotes extrapolated values based on the reported results.
			\end{tablenotes}
		\end{threeparttable}
		
	\end{table}

	\quad For image annotation tasks, AVeCQ~\cite{koutsos2024mathsf} reports an on-chain cost of 19 MWei for 39 workers. The cost of average review tasks scales approximately linearly with the number of workers, reaching 55 MWei for 128 workers. Thus, AVeCQ has a cost that is proportional to task complexity and worker participation. Duaan et al.~\cite{duan2019aggregating} evaluate on-chain review tasks. With 128 workers, the E2E latency is reported at approximately 30 minutes, and the on-chain cost reaches 416 MWei per task, mainly due to complete on-chain execution.
	
	\quad $\mathsf{DARTIC}$ outperforms both frameworks in terms of E2E latency. With respect to on-chain costs, it outperforms Duaan et al.~\cite{duan2019aggregating} both with and without aggregation, and surpasses AVeCQ~\cite{koutsos2024mathsf} when aggregation is employed. This efficiency stems from offloaded computations, where task evaluation and threshold signature generation are performed off-chain (with latency $<$1s per submission), with only aggregated results submitted on-chain. As a result, $\mathsf{DARTIC}$ achieves task-agnostic cost efficiency and minimal latency.

	\subsection{Reputation Consistency and Integrity}
	
\begin{figure}[t]
		\centering
		\subfloat[Reputation Consistency]{
			\hspace{-0.22in}
			\includegraphics[width=0.48\columnwidth]{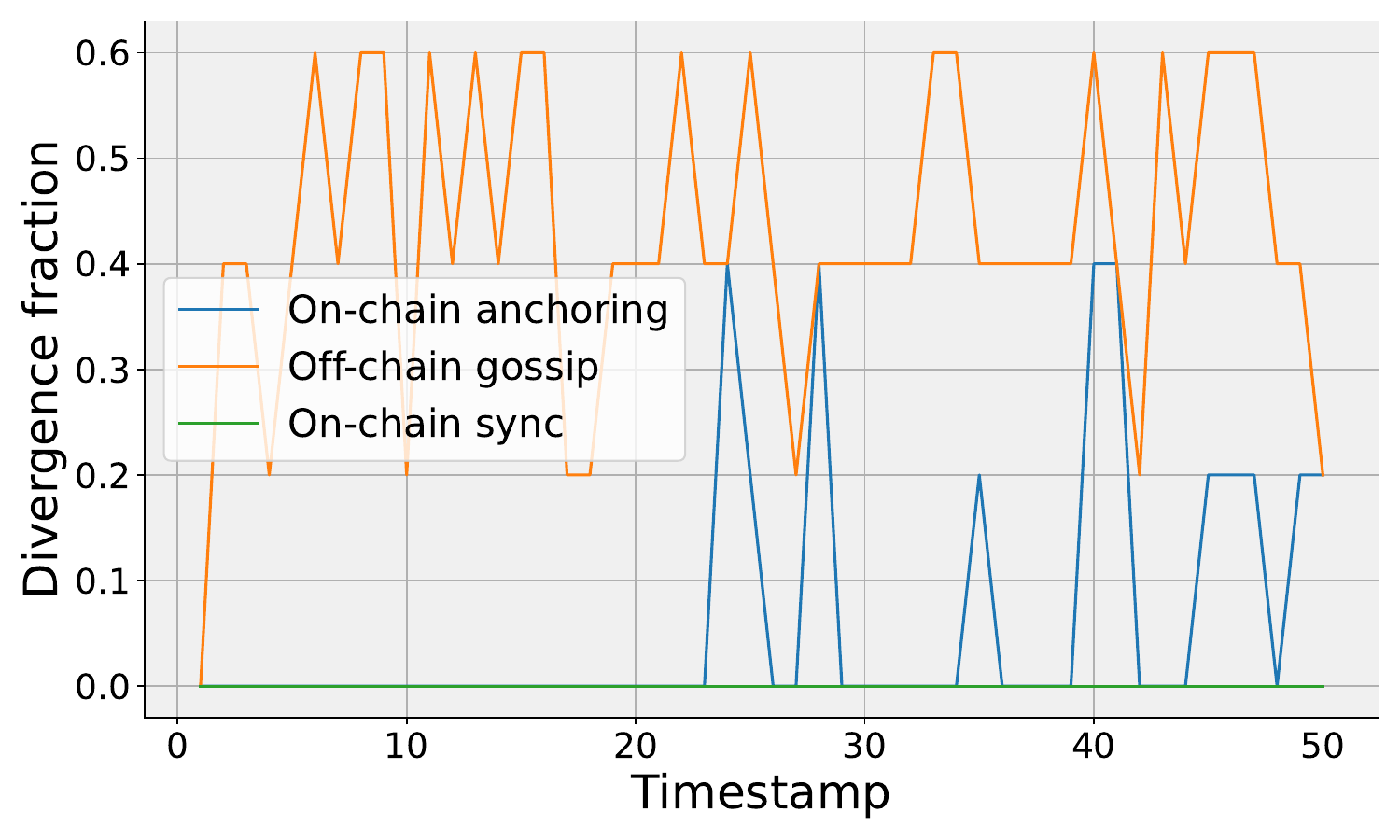}
			\label{fig:bc_role}
		}
		\subfloat[Reputation Integrity]{
			\hspace{-0.22in}
			\includegraphics[width=0.48\columnwidth]{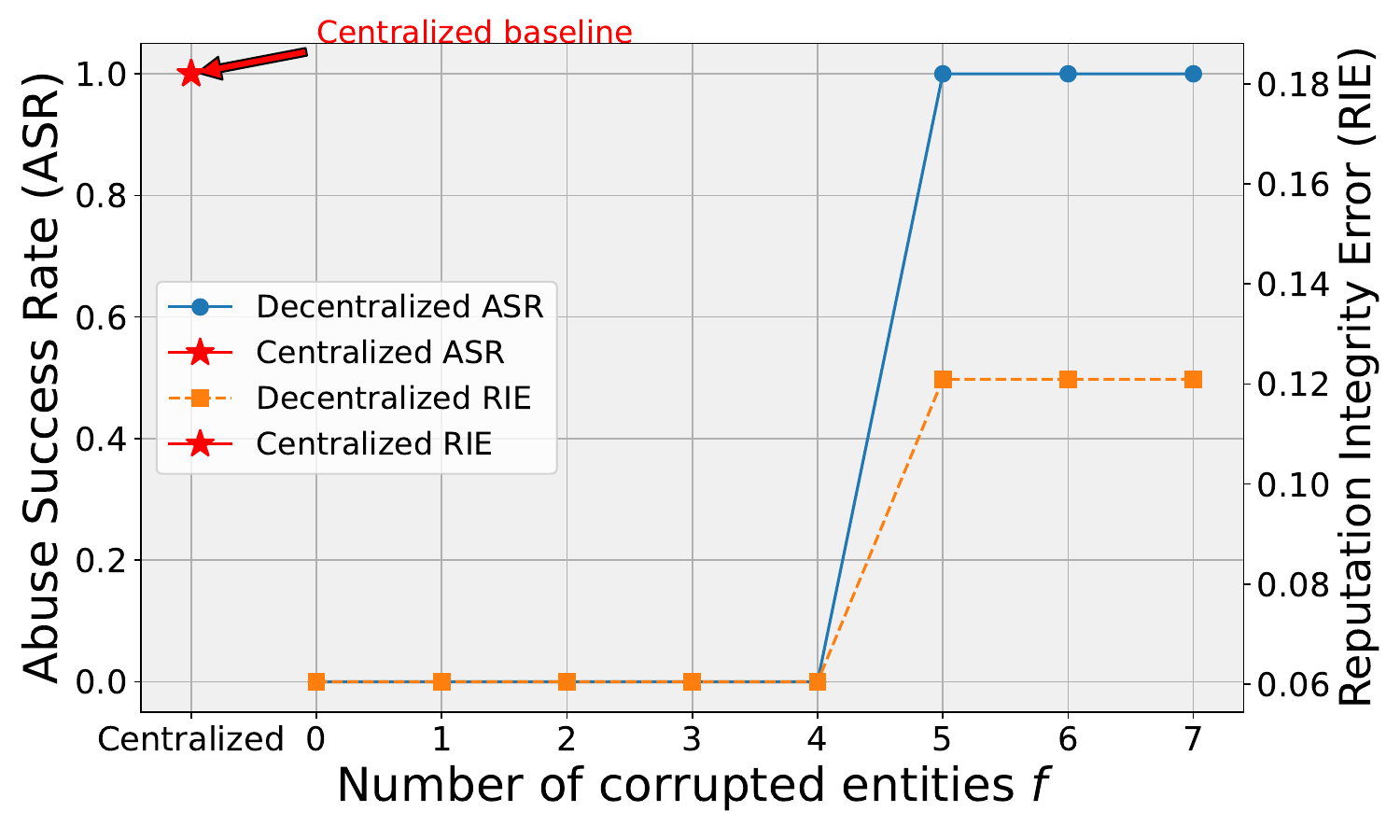}
			\label{fig:ts_role}
		}
		\caption{Comparison of reputation consistency and integrity.}
	\end{figure}

	To explicitly evaluate the functional role of blockchain in DARTIC, we conduct two complementary experiments. \textit{Exp~A} isolates the impact of on-chain anchoring on reputation consistency, while \textit{Exp~B} evaluates whether decentralized threshold-signature authorities prevent manipulation of reputation states. Both experiments compare the proposed design against counterfactual off-chain alternatives (\eg, centralized~\cite{anoTFPP}, P2P gossip~\cite{drep}). Full experimental setup, parameters and threat model are presented in Appendix~C.
	\paragraph*{Exp~A: Reputation State Consistency} We simulate a distributed ecosystem of mutually distrustful service raters, each maintaining local reputations. Updates are propagated either via probabilistic gossip (off-chain) or on-chain anchoring. Fig.~\ref{fig:bc_role} shows that on-chain mechanisms maintain tightly clustered reputations, whereas off-chain gossip allows transient divergence. This shows that on-chain anchoring with rapid finality is essential for consistent reputation states across raters.
	\paragraph*{Exp~B: Authority Abuse Resistance} We study threshold ECDSA enforcement for protecting reputation integrity against malicious or compromised oracles. A reputation update is finalized on-chain only if a quorum of authorities signs it. \begin{inparaenum}[(i)]
		\item \textit{Centralized authority vulnerability:} Even a single compromised oracle can arbitrarily manipulate reputations.
		\item \textit{Threshold resilience:} Threshold enforcement preserves integrity up to the corruption threshold ($t = \lfloor 2n/3 \rfloor + 1$), limiting deviation from honest reputation trajectories.
		\item \textit{Decentralized robustness:} ASR and RIE remain negligible until a quorum is corrupted, confirming that threshold signatures are necessary to resist unilateral or collusive authority abuse.\end{inparaenum}

	\section{Conclusion} \label{sec:$DARTC$conclusion}
	
	\quad This work presented how on-chain crowdsourcing can reconcile unlinkable anonymity, robust reputation binding, and scalability without relying on trusted intermediaries. We introduced $\mathsf{DARTIC}$, a decentralized and anonymous trust framework that decouples identity management from service execution through a dual-ledger architecture and zkSNARK set membership proofs. This design enables users to operate under multiple unlinkable pseudonyms while preventing Sybil and reputation-reset attacks. In addition, we designed an automated, privacy-preserving reputation model driven by verifiable service outcomes to maintain accountability. For practical deployability, we studied proof aggregation and zk-compatible batching mechanisms to significantly reduce verification overhead and on-chain costs. Experimental evaluation demonstrates that $\mathsf{DARTIC}$ achieves sub-3s proof generation, sub-second on-chain verification, and over an order-of-magnitude improvement in verification efficiency. These results confirm $\mathsf{DARTIC}$’s capability to support large-scale, anonymous, trustless Web3 crowdsourcing with strong resistance to identity and reputation manipulation.

	\bibliographystyle{IEEEtran}
	\bibliography{ref}

\end{document}